\documentclass[vecphys]{svmult}

\usepackage{makeidx}        
\usepackage{graphicx}        
                            
\usepackage{multicol}
\usepackage[bottom]{footmisc}

\usepackage[breaklinks]{hyperref}
\usepackage{natbib}

\newcommand{\bea}{\begin{eqnarray}}
\newcommand{\eea}{\end{eqnarray}}
\newcommand{\bc}{\begin{center}}
\newcommand{\ec}{\end{center}}

\renewcommand{\vec}[1]{ {\bf #1} }

\makeindex        
\begin{document}

\title*{Hydrodynamic simulations on a moving Voronoi mesh}
\author{Volker Springel}
\institute{Heidelberg Institute for Theoretical Studies,
  Schloss-Wolfsbrunnenweg 35, 69118 Heidelberg, Germany, \texttt{volker.springel@h-its.org}\vspace*{0.2cm}\\
Zentrum f\"ur Astronomie der Universit\"at Heidelberg, ARI,
  M\"{o}nchhofstr. 12-14, 69120 Heidelberg, Germany, \texttt{volker.springel@uni-hd.de}}
 
\maketitle

{\bf Summary} At the heart of any method for computational fluid
dynamics lies the question of how the simulated fluid should be
discretized.  Traditionally, a fixed Eulerian mesh is often employed
for this purpose, which in modern schemes may also be adaptively
refined during a calculation. Particle-based methods on the other hand
discretize the mass instead of the volume, yielding an approximately
Lagrangian approach. It is also possible to achieve Lagrangian
behavior in mesh-based methods if the mesh is allowed to move with the
flow.  However, such approaches have often been fraught with
substantial problems related to the development of irregularity in the
mesh topology. Here we describe a novel scheme that eliminates these
weaknesses. It is based on a moving unstructured mesh defined by the
Voronoi tessellation of a set of discrete points. The mesh is used to
solve the hyperbolic conservation laws of ideal hydrodynamics with a
finite volume approach, based on a second-order Godunov scheme with an
exact Riemann solver.  A particularly powerful feature of the approach
is that the mesh-generating points can in principle be moved
arbitrarily.  If they are given the velocity of the local flow, a
highly accurate Lagrangian formulation of continuum hydrodynamics is
obtained that is free of mesh distortion problems, while it is at the
same time fully Galilean-invariant, unlike ordinary Eulerian codes. We
describe the formulation and implementation of our new Voronoi-based
hydrodynamics, and we discuss a number of illustrative test problems
that highlight its performance in practical applications.

\section{Introduction}

Numerical simulations have become an indispensable tool to study fluid
dynamics, especially in astrophysics where direct experiments are
often impossible.  However, it is not always clear whether the
employed simulation algorithms are sufficiently accurate in real
practical applications, and to which extent numerical deficits may
affect the final results. It therefore remains an important task to
continue to critically test the numerical methods that are in use, and
to develop new approaches with the goal to reach better accuracy at
comparable or even lower computational cost.

In astrophysics, a variety of fundamentally quite different numerical
methods for hydrodynamical simulations are in use, the most prominent
ones are smoothed particle hydrodynamics
\citep[SPH;][]{Lucy1977,Gingold1977,Monaghan1992,Springel2010b} and
Eulerian mesh-based hydrodynamics
\citep[e.g.][]{Toro1997,LeVeque2002,Stone2008} with (optional)
adaptive mesh refinement (AMR). A particular challenge in astronomy is
the need to calculate self-gravitating flows, which often tend to
cluster strongly under gravity, producing a huge dynamic range in
density and length scales that can only be treated efficiently with
spatially adaptive resolution.  An important reason for the popularity
of SPH lies in the fact that such an adaptivity is automatically built
into this method, whereas achieving it in adaptive mesh refinement
codes requires substantial effort.

It has become clear over recent years that both SPH and AMR suffer
from fundamental problems that make them inaccurate in certain
regimes.  Indeed, these methods sometimes yield conflicting results
even for basic calculations that only consider non-radiative
hydrodynamics \citep[e.g.][]{Frenk1999, Agertz2007, Tasker2008,
  Mitchell2008}.  SPH codes have comparatively poor shock resolution,
offer only low-order accuracy for the treatment of contact
discontinuities, and suffer from subsonic velocity noise
\citep{Abel2011}. Worse, they appear to suppress fluid instabilities
under certain conditions \citep{Agertz2007}, as a result of a spurious
surface tension and inaccurate gradient estimates across density
jumps. On the other hand, Eulerian codes are not free of fundamental
problems either.  They do not produce Galilean-invariant results,
which can make their accuracy sensitive to the presence of bulk
velocities \citep[e.g.][]{Wadsley2008,Tasker2008}.  Another concern
lies in the mixing inherent in multi-dimensional Eulerian
hydrodynamics. This provides for an implicit source of entropy, with
sometimes unclear consequences \citep[e.g.][]{Wadsley2008}.

There is hence substantial motivation to search for new hydrodynamical
methods that improve on these weaknesses of the SPH and AMR
techniques. In particular, we would like to retain the accuracy of
mesh-based hydrodynamical methods (for which decades of experience
have been accumulated in computational fluid dynamics), while at the
same time we would like to outfit them with the Galilean-invariance
and geometric flexibility that is characteristic of SPH. The principal
idea for achieving such a synthesis is to allow the mesh to move with
the flow itself. This is an obvious and old idea \citep{Braun1995,
  Gnedin1995, Whitehurst1995, Mavripilis1997, Xu1997, Hassan1998,
  Pen1998, Trac2004}, but one fraught with many practical difficulties
that have so far prevented widespread use of any of the few past
attempts to introduce moving-mesh methods in astrophysics and
cosmology. For example, \citet{Gnedin1995} and \cite{Pen1998}
presented moving-mesh hydrodynamic algorithms which relied on the
continuous deformation of a Cartesian grid. However, the need to limit
the maximum allowed grid distortions severely impacts the flexibility
of these codes for situations in which the mesh becomes heavily
distorted, and special measures are required to let the codes evolve
cosmological density fields into a highly clustered state.  In
general, mesh tangling (manifested in `bow-tie' cells and hourglass
like mesh motions) is the traditional problem of such attempts to
simulate multi-dimensional hydrodynamics in a Lagrangian fashion.

\begin{figure}[t]
\centering
\resizebox{10cm}{!}{\includegraphics{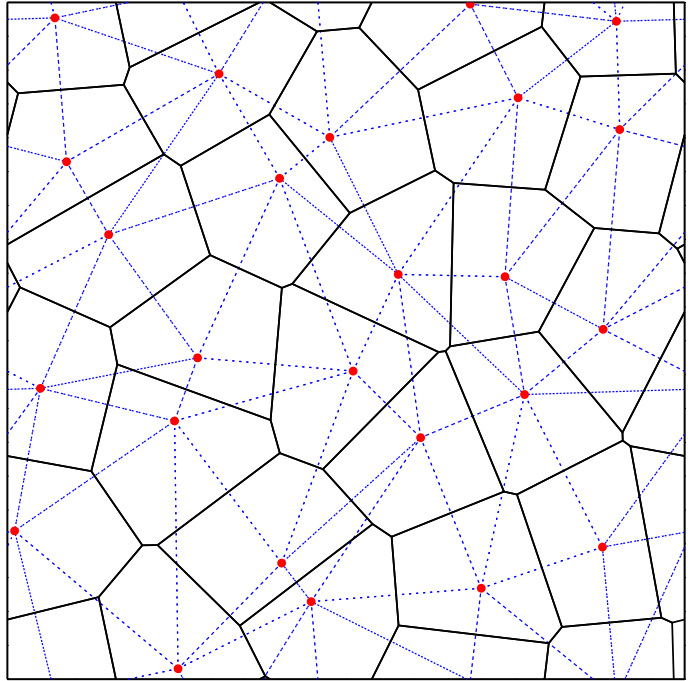}}
\caption{Example of a Voronoi and Delaunay tessellation in 2D, with periodic
  boundary conditions. The red circles show the generating points of the
  Voronoi tessellation, which is drawn with solid lines. Its topological dual,
  the Delaunay triangulation, is overlaid with thin dashed lines.
  \label{FigVoronoiExample}}
\end{figure}

In this contribution, we describe a new formulation of continuum
hydrodynamics based on an unstructured mesh. The mesh is defined as
the Voronoi tessellation of a set of discrete mesh-generating points,
which are in principle allowed to move freely.  For the given set of
points, the Voronoi tessellation of space consists of non-overlapping
cells around each of the sites such that each cell contains the region
of space closer to it than to any of the other sites.  Closely related
to the Voronoi tessellation is the Delaunay tessellation, the
topological dual of the Voronoi diagram. Both constructions have
already been widely used for natural neighbor interpolation and
geometric analysis of cosmic structures \citep[e.g.][]{Weygaert1994,
  Sambridge1995, Schaap2000, Pelupessy2003, vandeWeygaert2009}. In 2D,
the Delaunay tessellation for a given set of points is a triangulation
of the plane, where the points serve as vertices of the triangles.
The defining property of the Delaunay triangulation is that each
circumcircle around one of the triangles of the tessellation is not
allowed to contain any of the other mesh-generating points in its
interior. This empty circumcircle property distinguishes the Delaunay
triangulation from the many other triangulations of the plane that are
possible for the point set, and in fact uniquely determines the
triangulation for points in general position.  Similarly, in three
dimensions, the Delaunay tessellation is formed by tetrahedra that are
not allowed to contain any of the points inside their circumspheres.

As an example, Figure~\ref{FigVoronoiExample} shows the Delaunay and
Voronoi tessellations for a small set of points in 2D, enclosed in a
box with imposed periodic boundary conditions. The midpoints of the
circumcircles around each Delaunay triangle form the vertices of the
Voronoi cells, and for each line in the Delaunay diagram, there is an
orthogonal face in the Voronoi tessellation.

The Voronoi cells can be used as control volumes for a finite-volume
formulation of hydrodynamics, using the same principal ideas for
reconstruction, evolution and averaging (REA) steps that are commonly
employed in many Eulerian techniques. However, as we will see it is
possible to consistently include the mesh motion in the formulation of
the numerical steps, allowing the REA-scheme to become
Galilean-invariant.  Even more importantly, due to the mathematical
properties of the Voronoi tessellation, the mesh continuously deforms
and changes its topology as a result of the point motion, without ever
leading to the dreaded mesh-tangling effects that are the curse of
traditional moving mesh methods.  We note that the approach we
describe here is quite different from attempts to formulate fluid
particle models based on Voronoi cells \citep[e.g.][]{Hietel2000,
  Serrano2005, Hess2009}, or mesh-free finite volume approaches
\citep{Junk2002}. The former are similar in spirit to SPH and
typically maintain a constant mass per particle, whereas our scheme is
really closely related to ordinary mesh codes -- except that the mesh
is fully dynamic.
 
With illustrative test problems we shall later show that the resulting
formulation of hydrodynamics performs rather well on a number of test
problems, featuring very high accuracy in the treatment of shocks,
shear waves, and fluid instabilities. In particular, it can give
better results than fixed-mesh Eulerian hydrodynamics, thanks to much
reduced advection errors. It also offers much higher accuracy than SPH
when an equal number of particles/cells is used, making it highly
attractive as a possible alternative to currently employed SPH and AMR
schemes in astrophysics

This article is structured as follows.  In Section~\ref{SecHydro}, we
formulate continuum hydrodynamics on the Voronoi mesh, based on a
finite-volume ansatz and a second-order accurate extension of
Godunov's method. In Section~\ref{SecTime}, we briefly discuss time
integration and implementation aspects.  We then turn to a discussion
of a number of basic hydrodynamical tests in
Section~\ref{SecHydroTests}, chosen to highlight some of the principal
advantages and properties of the new approach. We note that a more
extensive discussion of code tests and of the algorithmic
implementation of the new scheme in the parallel {\small AREPO} code
can be found in \citet{Springel2010}. Finally, we summarize and
discuss our main findings in Section~\ref{SecDiscussion}.

\section{A finite volume discretization of the Euler equations on a moving Voronoi mesh} \label{SecHydro}

The Euler equations are conservation laws for mass, momentum and energy that
take the form of a system of hyperbolic partial differential equation. They
can be written in compact form by introducing a state vector
\begin{equation} 
\vec{U} = \left(
\begin{array}{c}
\rho\\
\rho \vec{v}\\
\rho e
\end{array} 
\right) = 
 \left(
\begin{array}{c}
\rho\\
\rho \vec{v}\\
\rho u + \frac{1}{2} \rho \vec{v}^2
\end{array} 
\right) 
\end{equation} 
for the fluid, where $\rho$ is the mass density, $\vec{v}$ is the
velocity field, and $e= u + \vec{v}^2 / 2$ is the total energy per unit
mass. $u$ gives the thermal energy per unit mass, which for an ideal gas
is fully determined by the temperature. These fluid quantities are
functions of the spatial coordinates $\vec{x}$ and time $t$,
i.e.~$\vec{U}=\vec{U}(\vec{x},t)$, but for simplicity we will typically refrain from
explicitly stating this dependence in our notation.  Based on $\vec{U}$,
we can define a flux function
\begin{equation} 
\vec{F}(\vec{U}) = \left(
\begin{array}{c}
\rho\vec{v}\\
\rho \vec{v}\vec{v}^T + P\\
(\rho e + P)\vec{v}
\end{array}
\right) ,
\end{equation} 
with an equation of state
\begin{equation} 
P= (\gamma-1)\rho u 
\end{equation}  
that gives the pressure of the fluid.  The Euler equations can then be
written in the compact form
\begin{equation}
\frac{\partial\vec{U}}{\partial t} +
\vec{\nabla}\cdot \vec{F} = 0,  \label{EqnEuler}
\end{equation}  
which emphasizes their character as conservation laws for mass, momentum
and energy.

Over the past decades, a large variety of different numerical
approaches to solve this coupled set of partial differential equations
have been developed \citep[see][for comprehensive
expositions]{Toro1997,LeVeque2002}.  We will here employ a so-called
{\em finite-volume} strategy, in which the discretization is carried
out in terms of a subdivision of the system's volume into a finite
number of disjoint cells. The fluid's state is described by the
cell-averages of the conserved quantities for these cells. In
particular, integrating the fluid over the volume $V_i$ of cell $i$,
we can define the total mass $m_i$, momentum $p_i$ and energy $E_i$
contained in the cell as follows,
\begin{equation}
\vec{Q}_i = \left(
\begin{array}{c}
m_i\\
\vec{p}_i\\
E_i 
\end{array}
\right)
= \int_{V_i} \vec{U}\,{\rm d} V .
\end{equation} 
With the help of the Euler equations, we can calculate the rate of
change of $\vec{Q}_i$ in time. Converting the volume integral over the
flux divergence into a surface integral over the cell results in
\begin{equation}
\frac{{\rm d}\vec{Q}_i}{{\rm d}t}
= -\int_{\partial V_i} \left[ \vec{F}(\vec{U}) - \vec{U}
\vec{w}^T\right] {\rm d}\vec{n} .
\label{EqQevol}
\end{equation} 
Here $\vec{n}$ is an outward normal vector of the cell surface, and
$\vec{w}$ is the velocity with which each point of the boundary of the
cell moves. In Eulerian codes, the mesh is taken to be static, so that
$\vec{w}=0$, while in a fully Lagrangian approach, the surface would
move at every point with the local flow velocity,
i.e.~$\vec{w}=\vec{v}$. In this case, the right hand side of
equation~(\ref{EqQevol}) formally simplifies, because then the first
component of $\vec{Q}_i$, the mass, stays fixed for each
cell. Unfortunately, it is normally not possible to follow the
distortions of the shapes of fluid volumes exactly in
multi-dimensional flows for a reasonably long time, or in other words,
one cannot guarantee the condition $\vec{w}=\vec{v}$ over the entire
surface. In this case, one needs to use the general formula of
equation~(\ref{EqQevol}), as we will do in this work.

The cells of our finite volume discretization are polyhedra with flat
polygonal faces (or lines in 2D). Let $\vec{A}_{ij}$ describe the
oriented area of the face between cells $i$ and $j$ (pointing from $i$
to $j$).  Then we can define the averaged flux across the face $i$-$j$ as
\begin{equation}
\vec{{F}}_{ij} = \frac{1}{A_{ij}} \int_{A_{ij}} \left[ \vec{F}(\vec{U})
- \vec{U} \vec{w}^T\right] {\rm d}\vec{A}_{ij},
\end{equation}  
and the Euler equations in finite-volume form become
\begin{equation} 
\frac{{\rm d}\vec{Q}_i}{{\rm d}t} = - \sum_j A_{ij} \vec{F}_{ij}.
\end{equation}  
We obtain a manifestly conservative time discretization of this equation
by writing it as
\begin{equation} 
\vec{Q}_i^{(n+1)} = \vec{Q}_i^{(n)} - \Delta t \sum_j A_{ij}
\vec{{\hat F}}_{ij}^{(n+1/2)}, 
\label{eqnupdate}
\end{equation} 
where the  $\vec{{\hat F}}_{ij}$ are now an appropriately time-averaged
approximation to the true flux $\vec{F}_{ij}$ across the cell face.  The
notation $\vec{Q}_i^{(n)}$ is meant to describe the state of the system
at step $n$.  Note that $\vec{{\hat F}}_{ij}$ = $-\vec{{\hat F}}_{ji}$,
i.e.~the discretization is manifestly conservative.

Evidently, a crucial step lies in obtaining a numerical estimate of the
fluxes $\vec{{\hat F}}_{ij}$, and a good fraction of the literature on
computational fluid dynamics is concerned with this problem. This issue
is particularly important since the most straightforward (and perhaps
naive) approach for estimating the fluxes, namely simply approximating
them as the average of the left and right cell-centered fluxes
catastrophically fails and invariably leads to severe numerical
integration instabilities that render such a scheme completely useless
in practice.

\begin{figure}[t]
\centering
\resizebox{11cm}{!}{\includegraphics{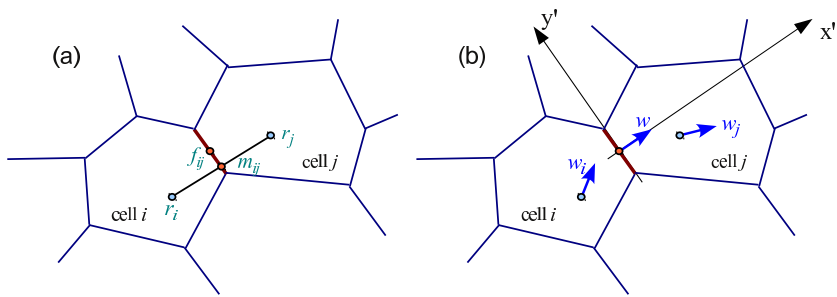}}
\caption{Sketch of a Voronoi mesh and the relevant geometric quantities that
  enter the flux calculation across a face. In (a), we show the the
  mesh-generating points $\vec{r}_{i}$ and $\vec{r}_{j}$ of two cells $i$ and
  $j$. The face between these two cells has a center-of-mass vector
  $\vec{f}_{ij}$, which in general will be offset from the mid-point $m_{ij}$
  of the two points. In (b), we illustrate the two velocity vectors $\vec{w}_{i}$
  and $\vec{w}_{j}$ associated with the mesh-generating points. These are normally
  chosen equal to the gas velocity in the cells, but other choices are allowed
  too. The motion of the mesh-generating points uniquely determines the motion
  of the face between the cells. Only the normal velocity $\vec{w}$ is however
  needed for the flux computation in the rotated frame $x'$, $y'$.
  \label{FigSketch}}
\end{figure}
 
Many modern schemes for estimating the fluxes in a stable fashion are
descendants of Godunov's method, which revolutionized the field. By
solving an exact or approximate Riemann problem at cell boundaries,
Godunov's method allows the correct identification of the
eigenstructure of the local solution and of the upwind direction,
which is crucial for numerical stability. While Godunov's original
method offers only first order accuracy and is relatively diffusive,
it can be extended to higher-order accuracy relatively simply, and in
many different ways.  We employ the MUSCL-Hancock scheme
\citep{Leer1984,Toro1997,Leer2006}, which is a well-known and
relatively simple approach for obtaining second-order accuracy in
space and time. This scheme is also popular in astronomy and used in
several state-of-the art Eulerian codes
\citep[e.g.][]{Fromang2006,Mignone2007,Cunningham2007}. In its basic
form, the MUSCL-Hancock scheme involves a slope-limited piece-wise
linear reconstruction step within each cell, a first order prediction
step for the evolution over half a timestep, and finally a Riemann
solver to estimate the time-averaged inter-cell fluxes for the
timestep. After the fluxes have been applied to each cell, a new
averaged state of the cells is constructed.  This sequence of steps in
a timestep hence follows the general REA approach.

Figure~\ref{FigSketch} gives a sketch of the geometry involved in
estimating the flux across the face between two Voronoi cells. Truly
multidimensional Riemann solvers have been developed recently
\citep{Wendroff1999,Brio2001,Balsara2010}, but it is unclear whether
they can be readily adapted to our complicated face geometry. We
therefore follow the common approach and calculate the flux for each
face separately, treating it as an effectively one-dimensional
problem. Since we do not work with Cartesian meshes, we cannot use
operating splitting to deal with the individual spatial
dimensions. Rather we use a method where all the fluxes are computed
in one step, and are then collectively applied to calculate the change
of the conserved quantities in a cell. This {\em unsplit} approach
implicitly accounts for ``corner fluxes'' \citep{Colella1990} needed
to recover second-order accuracy through the half-step prediction step
in our MUSCL-Hancock scheme (which exploits the primitive form of the
Euler equations).  For defining the Riemann problem normal to a cell
face, we rotate the fluid state into a suitable coordinate system with
the $x'$-axis normal to the cell face (see sketch).  This defines the
left and right states across the face, which we pass to an exact
Riemann solver. The latter is implemented following \citet{Toro1997}
with an extension to treat vacuum states, but it could easily be
substituted with an approximate Riemann solver for higher performance,
if desired.  We note that in multi dimensions the transverse
velocities are also required in the Riemann problem in order to
identify the correct upwind transverse velocity, which is important
for an accurate treatment of shear.  Once the flux has been calculated
with the Riemann solver, we transform it back to the lab frame.

A further important point concerns the treatment of the allowed motion
of cell surfaces in our scheme. In order to obtain stable upwind
behavior, the Riemann problem needs to be solved {\em in the frame of
  the moving face}. This is important as the one-dimensional Riemann
problem is not Galilean-invariant in the following sense: Suppose left
and right state at an interface are described by $(\rho_{\rm L}, P_{\rm
  L}, v_{\rm L})$ and $(\rho_{\rm R}, P_{\rm R}, v_{\rm R})$, for which
the Riemann solver returns an interface state $(\rho_{\rm F}, P_{\rm F},
v_{\rm F})$ that is the basis for the flux estimate. For example, the
mass flux across the interface is then given by $\rho_{\rm F} v_{\rm
  F}$. Consider now a velocity boost $v$ applied both to the left and
the right side. The new Riemann problem is given by $(\rho_{\rm L},
P_{\rm L}, v_{\rm L}+v)$ and $(\rho_{\rm R}, P_{\rm R}, v_{\rm R}+v)$,
and will return a flux estimate $\rho'_F v'_F$. However, in general this
will yield $\rho'_{\rm F} v'_{\rm F} \ne \rho_{\rm F}(v_{\rm F}+v)$,
which implies that the calculated flux vector is not Galilean
invariant.

In our new hydrodynamical scheme, each timestep involves the following
basic steps:
\begin{enumerate}
\item Calculate a new Voronoi tessellation based on the current coordinates
  $\vec{r}_i$ of the mesh generating points. This also gives the
  centers-of-mass $\vec{s}_i$ of each cell, their volumes $V_i$, as well as the
  areas $A_{ij}$ and centers $\vec{f}_{ij}$ of all faces between cells.
\item Based on the vector of conserved fluid variables $\vec{Q}_i$ associated
  with each cell, calculate the `primitive' fluid variables
  $\vec{W}_i=(\rho_i, \vec{v}_i, P_i)$ for each cell.
\item Estimate the gradients of the density, of each of the velocity components,
  and of the pressure in each cell, and apply a slope-limiting procedure to
  avoid overshoots and the introduction of new extrema.
\item Assign velocities $\vec{w}_i$ to the mesh generating points.
\item Evaluate the Courant criterion and determine a suitable timestep size
  $\Delta t$.
\item For each Voronoi face, compute the flux $\vec{{\hat F}}_{ij}$ across it
  by first determining the left and right states at the midpoint of the face
  by linear extrapolation from the cell midpoints, and by predicting these
  states forward in time by half a timestep. Solve the Riemann problem in a
  rotated frame that is moving with the speed of the face, and transform the
  result back into the lab-frame.
\item For each cell, update its conserved quantities with the total flux over
  its surface multiplied by the timestep, using equation
  (\ref{eqnupdate}). This yields the new state vectors $\vec{Q}_i^{(n+1)}$ of
  the conserved variables at the end of the timestep.
\item Move the mesh-generating points with their assigned velocities for this
  timestep.
\end{enumerate}
For the sake of definiteness, we now briefly describe the most important
details of these different steps.

\subsection{Gradient estimation and linear reconstruction} \label{SecGradients}

According to the Green-Gauss theorem, the surface integral of a scalar
function over a closed volume is equal to its gradient integrated over
the same volume, i.e.
\begin{equation}
\int_{\partial V} \phi\,  {\rm d} \vec{n} =
\int_{V} \vec{\nabla}\phi \,{\rm d}V.
\end{equation} 
This suggests one possible way to estimate the mean gradient in a Voronoi cell,
in the form
\begin{equation}
\left<\vec{\nabla}\phi\right>_i
\simeq
-\frac{1}{V_i}\sum_j
 \phi(\vec{f}_{ij}) \,\vec{A}_{ij},
\label{Eqnbasicgrad}
\end{equation}
where $\phi(\vec{f}_{ij})$ is the value of $\phi$ at the centroid
$\vec{f}_{ij}$ of the face shared by cells $i$ and $j$, and $\vec{A}_{ij}$ is
a vector normal to the face (from $j$ to $i$), with length equal to the face's
area. Based on the further approximation
\begin{equation}
 \phi(\vec{f}_{ij})\simeq \frac{1}{2}(\phi_i + \phi_j),
\label{eqnmidp}
\end{equation}
this provides an estimate for the local gradient. Note that with the use
of equation (\ref{eqnmidp}), the gradient of cell $i$ only depends on
the values $\phi_j$ of neighboring cells, but not on $\phi_i$
itself. While the estimate~(\ref{Eqnbasicgrad}) can be quite generally
applied to arbitrary tessellations, due to the use of only one Gauss
point per face it is also relatively inaccurate and is not exact to
linear order in general.

For the special case of Voronoi cells, it is however possible to obtain a
considerably better gradient estimate with little additional effort. The key
is to carry out the surface integral more accurately. It can be shown
\citep{Serrano2001,Springel2010} that the gradient estimate
\begin{equation}
\left<\vec{\nabla}\phi\right>_i =   \frac{1}{V_i}
   \sum_{j\ne i}  A_{ij} \left( [\phi_j -\phi_i]\,  \frac{\vec{c}_{ij}}{r_{ij}}  
-\frac{\phi_i + \phi_j}{2}\,\frac{\vec{r}_{ij}}{r_{ij}} \right)
\label{eqngrad2}
\end{equation}
is exact to linear order, independent of the locations of the mesh-generating
points of the Voronoi tessellation.  Here we followed the notation of
\citet{Serrano2001} and defined $\vec{c}_{ij}$ as the vector from the midpoint
between $i$ and $j$ to the center-of-mass of the face between $i$ and $j$.
Without the term involving $\vec{c}_{ij}$ this gradient estimate is the same as
the simpler Green-Gauss estimate. However, retaining this extra term leads to
significantly better accuracy, because the gradient estimate becomes exact to
linear order for arbitrary Voronoi meshes. In practice, we shall therefore
always use this gradient estimation in our MUSCL-Hancock scheme for the Euler
equations, where we calculate in this way gradients for the 5 primitive
variables $(\rho, v_x, v_y, v_z, P)$ that characterize each cell.

The result (\ref{eqngrad2}) has also an interesting relation to the
formulae obtained by \citet{Serrano2001} for the partial derivatives of
the volume of a Voronoi cell with respect to the location of one of the
points.  As \citet{Serrano2001} have shown, the derivative of the volume
of a Voronoi cell due to the motion of a surrounding point is given by
\begin{equation}
\frac{\partial V_i}{\partial \vec{r}_j}
= -A_{ij} \left(\frac{\vec{c}_{ij}}{r_{ij}}  
+ \frac{\vec{r}_{ij}}{2r_{ij}} \right) \;\;\; {\rm for}\;\;\; i\ne j .
\label{EqSerr1}
\end{equation}
Furthermore, they show that
\begin{equation}
  \frac{\partial V_i}{\partial \vec{r}_i}
  = - \sum_{j\ne i} \frac{\partial V_j}{\partial \vec{r}_i}.
\label{EqSerr2}
\end{equation}
Using these relations, and noting that according to the Gauss theorem we have
\begin{equation}
 \frac{\phi_i}{V_i}   \sum_{j\ne i}  A_{ij} \frac{\vec{r}_{ij}}{r_{ij}} = 0 ,
\end{equation}
because the summation is just the surface integral of a constant function,
we can also write the estimate for the gradient 
 of $\phi$ at $\vec{r}_i$ more
compactly as
\begin{equation}
\left<\vec{\nabla}\phi\right>_i = -\frac{1}{V_i}\sum_j \frac{\partial V_j}{\partial
  \vec{r}_i} \phi_j .
\end{equation}
An interesting corollary of the above is that
provided $\phi(\vec{r})$ varies only linearly, 
the sum
\begin{equation}
S = \sum_i \phi(\vec{r}_i) V_i
\end{equation}
approximates the integral $\int \phi(\vec{r}) \,{\rm d}V$ {\em exactly},
independent of the positions of the points that generate the Voronoi
tessellation.

In our approach, we use the gradients estimated with equation (\ref{eqngrad2})
for a linear reconstruction in each cell around the center-of-mass. For
example, the density at any point $\vec{r}\in V_i$ of a cell is estimated
as
\begin{equation} \rho(\vec{r}) = \rho_i +
\left<\vec{\nabla}\rho\right>_i \cdot(\vec{r} - \vec{s}_i), 
\end{equation} 
where $\vec{s}_i$ is the center of mass of the cell. Note that independent of
the magnitude of the gradient and the geometry of the Voronoi cell, this
linear reconstruction is conservative, i.e.~the total mass in the cell $m_i$
is identical to the volume integral over the reconstruction, $m_i =
\int_{V_i}\rho( \vec{r}){\rm d}^3 r$. An alternative choice for the reference
point is to choose the mesh-generating point $\vec{r}_i$ instead of
$\vec{s}_i$. This is the more natural choice if the cell values are known to
sample the values of the underlying field at the location of the
mesh-generating points, then the reconstruction is exact to linear order.
However, our input quantities are cell-averages, which correspond to linear
order to the values of the underlying field sampled at the center-of-masses of
the cells. For this reason we prefer the center-of-mass of a cell as reference
point for the reconstruction.

Nevertheless, this highlights that large spatial offsets between the
center-of-mass of a cell and its mesh-generating point are a source of errors
in the linear reconstruction.  It is therefore desirably to use ``regular''
meshes if possible, where the mesh-generating points lie close to the
center-of-mass; such meshes minimize the errors in the gradient estimation and
the linear reconstruction. Or in other words, we would like our Voronoi meshes
to be relatively close to so-called {\em centroidal Voronoi meshes}, where the
mesh-generating points lie exactly in the center of mass of each cell. As we
discuss in bit more detail later, we have developed an efficient method for
steering the mesh motion such that this regularity condition can be
approximately maintained at all times.

\subsection{Slope limiting procedure}

In smooth parts of the flow, the above reconstruction is second-order
accurate. However, in order to avoid numerical instabilities the order of the
reconstruction must be reduced near fluid discontinuities, such that the
introduction of new extrema by over- or undershoots in the extrapolation is
avoided.  This is generally achieved by applying slope limiters that reduce
the size of the gradients near local extrema, or by flux limiters that replace
the high-order flux with a lower order version if there are steep gradients in
the upstream region of the flow.

We here generalize the original MUSCL approach to an unstructured grid by
enforcing monotonicity with a slope limiting of the gradients. To this end we
require that the linearly reconstructed quantities on face centroids do not
exceed the maxima or minima among all neighboring cells \citep{Barth1989}.
Mathematically, we replace the gradient with a slope-limited gradient
\begin{equation}
\left<\vec{\nabla}\phi\right>_i^{'} = \alpha_i
\left<\vec{\nabla}\phi\right>_i ,
\end{equation}
where the slope limiter $0\le \alpha_i\le 1$ for each cell is computed as 
\begin{equation}
\alpha_i = \min(1, \psi_{ij}).
\end{equation}
Here the minimum is taken with respect to all cells $j$ that are
neighbors
of cell $i$, and the quantity $\psi_{ij}$ is defined as
\begin{equation}
\psi_{ij} = \left\{ 
\begin{array}{ccc}
(\phi_i^{\rm max} -  \phi_i)/ \Delta\phi_{ij}  & {\rm for} & \Delta\phi_{ij}>0\\
(\phi_i^{\rm min} -  \phi_i)/ \Delta\phi_{ij}  & {\rm for} & \Delta\phi_{ij}<0\\
1 & {\rm for} & \Delta\phi_{ij}=0\\
\end{array}
\right.
\end{equation}
where $\Delta\phi_{ij}=\left<\vec{\nabla}\phi\right>_i\cdot(\vec{f}_{ij} -
\vec{s}_i)$ is the estimated change between the centroid $\vec{f}_{ij}$ and
the center of cell $i$, and $\phi_i^{\rm max} = \max(\phi_j)$ and $\phi_i^{\rm
  min} = \max(\phi_j)$ are the maximum and minimum values occurring for $\phi$
among all neighboring cells of cell $i$, including $i$ itself.
We note that this slope limiting scheme does not strictly enforce the total
variation diminishing property, which means that (usually reasonably small)
post-shock oscillations can sometimes still occur.

\subsection{Setting the velocities of the mesh generators}

A particular strength of the scheme we propose here is that it can be
used both as an Eulerian code, and as a Lagrangian scheme. The
difference lies only in the motion of the mesh-generation points. If
the mesh-generating points are arranged on a Cartesian mesh and zero
velocities are adopted for them, our method is identical to a
second-order accurate Eulerian code on a structured grid. Of course,
one can equally well choose a different layout of the points, in which
case we effectively obtain an Eulerian code on an unstructured
mesh. The real advantage of the new code can be realized when we allow
the mesh to move, with a velocity that is tied to the local fluid
speed. In this case, we obtain a Lagrangian hydrodynamical code, which
has some unique and important advantages relative to an Eulerian
treatment. It is however also possible to prescribe the mesh motion
through an external flow field, for example in order to smoothly
concentrate resolution towards particular regions of a mesh, or to
realize rotating meshes.  Unlike other arbitrary Lagrangian-Eulerian
(ALE) fluid dynamical methods, the method proposed here does however
not rely on remapping techniques to recover from distortions of the
mesh once they become severe, simply because the Voronoi tessellation
produced by the continuous motion of the mesh-generating points yields
a mesh geometry and topology that itself changes continuously in time,
without mesh-tangling effects.

The most simple and basic approach for specifying the motion of the mesh
generators is to use
\begin{equation}
\vec{w}_i = \vec{v}_i,
\label{eqnvelmesh}
\end{equation}
i.e.~the points are moved with the fluid speed of their cell. This
Lagrangian ansatz is clearly appropriate for pure advection and in
smooth parts of the flow. However, in this scheme there is no
mechanism built in that tries to improve the regularity of the Voronoi
mesh in case the mean mass per cell should develop substantial scatter
around a desired mean value, or if cells with high aspect ratios
occur. If desired, such tendencies of a growing mesh irregularity can
be counteracted by adding corrective velocity components to the
primary mesh velocities $\vec{w}_i$ of equation
(\ref{eqnvelmesh}). There are many different possibilities for how
exactly to do this, and we consider this freedom a strength of the
formalism. In Section~\ref{SecMeshRegularity}, we will discuss a
simple regularization method that we have found to be very effective.

\subsection{Flux computation}

An important aspect of our approach is that the specified velocities of the
mesh-generating points fully determine the motion of the whole Voronoi mesh,
including, in particular, the velocities of the centroids of cell faces (see
sketch in Fig.~\ref{FigSketch}).  This allows us to calculate the Riemann
problem in the rest-frame of each of the faces.

Consider one of the faces in the tessellation and call the fluid states in the
two adjacent cells the `left' and `right' states. We first need to determine
the velocity $\vec{w}$ of the face based on the velocities $\vec{w}_{\rm L}$
and $\vec{w}_{\rm R}$ of the two mesh-generating points associated with the
face (they are connected by a Delaunay edge). It is clear that $\vec{w}$ has a
primary contribution from the mean velocity $(\vec{w}_{\rm L}+\vec{w}_{\rm
  R})/2$ of the points, but there is also a secondary contribution $\vec{w}'$
from the residual motion of the two points relative to their center of
mass. This residual motion is given by $\vec{w}_{\rm R}' = - \vec{w}_{\rm L}'
= (\vec{w}_{\rm R} - \vec{w}_{\rm L})/2$, and we need to determine its impact
on the motion of the face centroid.  The components of $\vec{w}_{\rm R}'$ and
$\vec{w}_{\rm L}'$ parallel to the line connecting the centroid $\vec{f}$ of
the face with the midpoint $\vec{m}$ of the two mesh-generating points
$\vec{r}_{\rm L}$ and $\vec{r}_{\rm R}$ induce a rotation of the face around
the point $\vec{m}$. We are only interested in the normal velocity component
of this motion at the centroid  of the face. This can be easily computed as
\begin{equation}
\vec{w}' = \frac{(\vec{w}_{\rm L} - \vec{w}_{\rm R})
\cdot
[\vec{f}-(\vec{r}_{\rm R} + \vec{r}_{\rm L})/2]}{|\vec{r}_{\rm
    R}-\vec{r}_{\rm L}|}\,\,
\frac{(\vec{r}_{\rm R}-\vec{r}_{\rm L})}{|\vec{r}_{\rm R}-\vec{r}_{\rm L}|}.
\end{equation}
The full velocity $\vec{w}$ of the face is then given by
\begin{equation}
\vec{w} = \frac{\vec{w}_{\rm R} + \vec{w}_{\rm L}}{2} + \vec{w}'.
\end{equation}

We now calculate the flux across the face using the MUSCL-Hancock approach,
with the important difference that we shall carry out the calculation in the
rest-frame of the face. It is convenient to do this in the primitive variables
$(\rho, \vec{v}, P)$, where we first transform the lab-frame velocities of the
two cells to the rest-frame of the face by subtracting $\vec{w}$,
\begin{equation}
\vec{W}_{\rm L,R}' = \vec{W}_{\rm L,R} - \left(
\begin{array}{c}
0\\
\vec{w}\\
0\\
\end{array}\right) .
\label{eqnrefframe}
\end{equation}
We then linearly predict the states on both side to the centroid of the
face, and also predict them forward in time by half a timestep. This
produces the states 
\begin{equation}
\vec{W}_{\rm L,R}'' = \vec{W}_{\rm L,R}' + \left.\frac{\partial
    \vec{W}'}{\partial \vec{r}}\right|_{\rm L, R} (\vec{f} - \vec{s}_{\rm L,R})
+ \left.\frac{\partial
    \vec{W}'}{\partial t}\right|_{\rm L, R} \frac{\Delta t}{2} .
\end{equation}
The spatial derivatives $\partial \vec{W}'/\partial \vec{r}$ are known, and
given by the (slope-limited) gradients of the primitive variables that are
estimated as described in Section~\ref{SecGradients}. Note that the gradients are
unaffected by the change of rest-frame described by
Eqn.~(\ref{eqnrefframe}). The partial time derivate $\partial \vec{W}/\partial
t$ can be replaced by spatial derivatives as well, based on the Euler equations
in primitive variables, which are given by
\begin{equation}
\frac{\partial \vec{W}}{\partial t}
+ \vec{A}( \vec{W}) \frac{\partial \vec{W}}{\partial \vec{r}} = 0 ,
\end{equation}
where $\vec{A}$ is the matrix
\begin{equation}
\vec{A}( \vec{W}) = \left(
\begin{array}{ccc}
\vec{v} & \rho & 0 \\
 0     & \vec{v} & 1/\rho \\
0 & \gamma P & \vec{v}\\
\end{array}
\right).
\end{equation}
Having finally obtained the states left and right of the interface, we need to
turn them into a coordinate system aligned with the face, such that we can
solve an effectively one-dimensional Riemann problem. The required rotation
matrix $\vec{\Lambda}$ for the states only affects the velocity components,
viz.
\begin{equation}
\vec{W}_{\rm L,R}''' = \vec{\Lambda} \,\vec{W}_{\rm L,R}''=
 \left(
\begin{array}{ccc}
1 &  0 & 0 \\
 0  & \vec{\Lambda}_{\rm 3D} & 0 \\
0 &  0 & 1\\
\end{array}
\right)
\vec{W}_{\rm L,R}'',
\end{equation}
where $\vec{\Lambda}_{\rm 3D}$ is an ordinary rotation of the coordinate
system, such that the new $x$-axis is parallel to the normal vector of the
face, pointing from the left to the right state.

With these final states, we now solve the Riemann problem, and sample
the self-similar solution along $x/t=0$. This can be written as
\begin{equation}
\vec{W}_{\rm F} = R_{\rm iemann}(\vec{W}_{\rm L}''' , \vec{W}_{\rm R}'''),
\end{equation}
where $R_{\rm iemann}$ is a one-dimensional Riemann solver, which returns a
solution for the state of the fluid $\vec{W}_{\rm F}$ on the face in
primitive variables. We now transform this back to the lab-frame, reversing
the steps above,
\begin{equation}
\vec{W}_{\rm lab} = 
\left(
\begin{array}{c}
\rho\\
\vec{v}_{\rm lab}\\
P\\
\end{array}
\right) 
=
\Lambda^{-1}\vec{W}_{\rm F} + \left(
\begin{array}{c}
0\\
\vec{w}\\
0\\
\end{array}
\right) .
\end{equation}
Finally, we can use this state to calculate the fluxes in the conserved
variables across the face. Here we need to take into account that the face is
moving with velocity $\vec{w}$, meaning that the appropriate flux vector in
the lab frame is given by
\begin{equation}
\vec{{\hat F}} = \vec{F}(\vec{U}) - \vec{U}\vec{w}^{\rm T}=
\left(\begin{array}{c}
\rho(\vec{v}_{\rm lab}-\vec{w}) \\
\rho\vec{v}_{\rm lab}(\vec{v}_{\rm lab}-\vec{w})^{\rm T} + P\\
\rho e_{\rm lab} (\vec{v}_{\rm lab}-\vec{w})  + P\vec{v}_{\rm lab}
\end{array}
\right),
\label{eqfinalflux}
\end{equation}
where $\vec{U}$ is the state $\vec{W}_{\rm lab}$ expressed in the conserved
variables, and $e_{\rm lab}= \vec{v}^2_{\rm lab}/2 + P_{\rm lab}/[(\gamma-1)
\rho_{\rm lab}]$.  The scalar product of this flux vector with the normal
vector of the face gives the net flux of mass, momentum, and energy that the
two adjacent, moving cells exchange.  It is the flux of equation
(\ref{eqfinalflux}) that can finally be used in the conservative updates of
each cell, as described by equation~(\ref{eqnupdate}).

For the above formulation, it is straightforward to show that the changes in
the conserved quantities in a cell are Galilean invariant; any Galilean boost
is effectively simply absorbed into the motion of the face.

\section{Time integration and implementation aspects} \label{SecTime}

\subsection{Time integration}  \label{SecTimeintegration}

For hydrodynamics with a global timestep, we employ a simplified CFL
timestep criterion in the form
\begin{equation}
  \Delta t_i = C_{\rm CFL} \frac{ R_i}{c_i + |\vec{v}'_i|}  \label{EqnTiStep}
\end{equation}
to determine the maximum allowed timestep for a cell $i$. Here $R_i$ is the
effective radius of the cell, calculated as $R_i=(3V_i/4\pi)^{1/3}$ from the
volume of a cell (or as $R_i=(V_i/\pi)^{1/2}$ from the area in 2D), under the
simplifying assumption that the cell is spherical. The latter is normally a
good approximation, because we steer the mesh motion such that the
cell-generating point lies close to the center-of-mass of the cell, which
gives it a ``roundish'' polyhedral shape.  $C_{\rm CFL} < 1$ is the
Courant-Friedrichs-Levy coefficient (usually we choose $C_{\rm CFL}\simeq
0.4-0.8$), $c_i = \sqrt{\gamma P/\rho}$ is the sound speed in the cell, and $
|\vec{v}'_i| = |\vec{v}_i-\vec{w}_i|$ is the velocity of the gas {\em relative
  to the motion of the grid}. In the Lagrangian mode of the scheme, the
velocity $|\vec{v}'_i|$ is close to zero and usually negligible against the
sound speed, which means that larger timesteps than in an Eulerian treatment
are possible, especially if there are large bulk velocities in the system.

If the code is operated with a global timestep, we determine the next system
timestep as the minimum
\begin{equation}
\Delta t = \min_i \Delta t_i
\end{equation}
of the timestep limits of all particles.

It is also possible to implement an individual timestep scheme, where
the timestep conditions of different cells are treated in a more
flexible fashion. This can greatly improve the computational
efficiency in many applications. For example, in cosmological
simulations, a large dynamic range in densities quickly occurs as a
result of gravitational clustering. Accordingly, the local dynamical
times can vary by orders of magnitude. It has therefore long become
common practice to use individual timesteps for the collisionless
N-body problem, a technique that has also been extended to
hydrodynamical SPH simulations
\citep[e.g.][]{Katz1996,Springel2001gadget}. We have implemented such
a method also for the moving-mesh scheme, based on a discretization of
the allowed timestep sizes into a power-of-two hierarchy.  Unlike the
approach taken in AMR simulations, where refined grid patches are
typically as a whole subcycled in time by a constant factor, we impose
no such restriction on our mesh, i.e.~in principle each cell can be
evolved with its own timestep, constrained only to the power-of-two
hierarchy of allowed timestep sizes. To maintain a fully conservative
character of the scheme, we evolve each face with the smaller timestep
of the two adjacent cells. Full details of this individual timestep
scheme can be found in \citet{Springel2010}.

\subsection{Mesh regularity} \label{SecMeshRegularity}

As seen in Figure~\ref{FigVoronoiExample}, Voronoi meshes may
sometimes look quite ``irregular'', in the sense that there is a
significant spread in sizes and aspect ratios of the cells, especially
for disordered point distributions. While this is not a problem of
principle for our approach, it is clear that the computational
efficiency will normally be optimized if regions of similar gas
properties are represented with cells of comparable size.  Having a
mixture of cells of greatly different volumes to represent a gas of
constant density will restrict the size of the timestep unnecessarily
(which is determined by the smallest cells), without giving any
benefit in spatial resolution (which will be limited by the largest
cells in the region).

As we have seen, it is also desirable to have cells where the center-of-mass
lies close to the mesh-generating point, because this minimizes errors in the
linear reconstruction and limits the rate at which mesh faces turn their
orientation during mesh motion.  Below, we will discuss one possible approach
for steering the mesh motion during the dynamical evolution such that, if
desired, mesh regularity in the above sense can be achieved and maintained.

In so-called centroidal Voronoi tessellations \citep{Okabe2000}, the
mesh-generating points coincide with the center-of-mass of all
cells. There is an amazingly simple algorithm known as Lloyd's method
\citep{Lloyd1982} to obtain a centroidal Voronoi tessellation starting
from an arbitrary tessellation. One simply moves the mesh-generating
points of the current Voronoi tessellation to the center-of-masses of
their cells, and then reconstructs the Voronoi tessellation. The
process is repeated iteratively, and with each iteration, the mesh
relaxes more towards a configuration in which the Voronoi cells appear
quite `round' (in 2D they form a honeycomb-like mesh) and have similar
volume -- a centroidal Voronoi tessellation.

Inspired by this algorithm, we employ a simple scheme to improve, if
needed, the local shape of the Voronoi tessellation during the
dynamical evolution.  We simply augment equation (\ref{eqnvelmesh})
with an additional velocity component, which is designed to move a
given mesh-generating point towards the center-of-mass of its cell.
There are different possibilities to parameterize such a corrective
velocity.  One approach that we found to work quite well in practice is
to add a correction velocity whenever the mesh-generating point is
further away from the center-of-mass of a cell than a given threshold,
irrespective of the actual velocity field of the gas. To this end, we
associate a radius $R_i=(3V_i/4\pi)^{1/3}$ with a cell based on its
volume (or area in 2D). If the distance $d_i$ between the cell's
center-of-mass $\vec{s}_i$ and its mesh-generating point $\vec{r}_i$
exceeds some fraction $\eta$ of the cell radius $R_i$, we add a
corrective term proportional to the local sound speed $c_i$ of the
cell to the velocity of the mesh-generating point. This effectively
applies one Lloyd iteration (or a fraction of it) to the cell by
repositioning the mesh-generating point onto the current
center-of-mass, ignoring other components of the mesh motion.  In
order to soften the transition between no correction and the full
correction, we parameterize the velocity as
\begin{equation}
\vec{w}_i' = \vec{w}_i + \chi \left\{
\begin{array}{cl}
0 &   \mbox{for}\;d_i/(\eta\,R_i) <  0.9  \\
c_i \frac{\vec{s}_i-\vec{r}_i}{d_i}\frac{d_i-0.9\,\eta R_i}{0.2\,\eta R_i}  
& \mbox{for}\; 0.9 \le d_i/ (\eta\,R_i) <  1.1\\
 c_i\frac{\vec{s}_i-\vec{r}_i}{d_i}  & \mbox{for}\;1.1 \le d_i /(\eta\,R_i)
\end{array}
\right.
\label{EqnShapeCorrVel}
\end{equation}
but the detailed width of this transition is unimportant. In very cold
flows the sound speed may be so low that the correction becomes
ineffective. As an alternative, we therefore also implemented an option
in our code that allows a replacement of $c_s(\vec{s}_i-\vec{r}_i)/d_i$
in equation (\ref{EqnShapeCorrVel}) with
$({\vec{s}_i-\vec{r}_i})/{\Delta t}$. This more aggressive approach to
ensure round cells generally works very well too, but has the
disadvantage to depend on the timestepping.  Our typical choice for the
threshold of the correction is $\eta=0.25$, and we usually set
$\chi=1.0$, i.e.~the correction is, if present, applied in full over the
course of one timestep.  Smaller values of $\eta$ can be used to enforce
round cell shapes more aggressively, if desired. 

The above scheme is usually quite effective in maintaining low aspect
ratios and a regular mesh at all times during the evolution. However,
the criterion for detecting cells that should get a correction
velocity is not triggered if a mesh is strongly stretched or
compressed in one direction; then the centers of mass of cells can
still be close to their mesh-generating points, but the aspect ratio
of cells can be very high. We have found \citep{Vogelsberger2011} that
a simple alternative criterion treats such situations much better. To
this end we determine for each cell the maximum angle under which any
of the faces of the cell is seen from its mesh-generating point. If
this angle lies above a prescribed threshold value, the
mesh-correction component to the velocity is invoked, just as
above. This approach will effectively try to prevent that a
mesh-generating point gets too close to an outer wall of a cell, which
simultaneously ensures that the displacement from the center-of-mass
and the aspect ratio stay small. We have also found that with this
criterion the mesh-correction motions are required more rarely, hence
we have made this our default choice for general simulations with the
moving mesh approach. In any case, it is important to note that the
correction velocities are still Galilean-invariant, and they vanish
most of the time, so that the mesh-generating points will usually be
strictly advected with the local fluid velocity.

We point out that there is an important difference of this approach
compared with the mesh regularization technique presented in
\citet{Hess2009}.  In the finite volume approach discussed here, one
may in principle move the mesh-generating points in nearly arbitrary
ways. Maintaining a good mesh is therefore comparatively
straightforward, as described above. In contrast, the Voronoi particle
model of \citet{Hess2009} dictates a particular equation-of-motion for
the mesh-generating points, where one is not allowed to simply add
some mesh correction velocities. As a way out, \citet{Hess2009}
suggested to modify the underlying Lagrangian in a tricky way in order
to automatically build in corrective motions into the dynamics of the
mesh-generating points, but this approach is not equally flexible as
the one we can use here.

\subsection{Implementation aspects}

The scheme described thus far has been implemented in the {\small AREPO} code,
which is described in detail in \citet{Springel2010}. A central aspect of the
code is a fast engine for the generation of Delaunay and Voronoi meshes. To
this end an incremental insertion algorithm is used both in 2D and 3D, which
also allows partial mesh constructions, as needed in our individual timestep
approach. For reasons of memory and run-time efficiency, we have written our
own low-level mesh-construction routines instead of using a library such as
{\small CGAL}. The mesh construction is parallelized for distributed memory
machines. We use a spatial domain decomposition in which each domain is mapped
to a single processor, which then first constructs its part of the mesh
independently of the other CPUs, and then exchanges and inserts additional
`ghost' particles as needed to make sure that the Voronoi cells of all local
particles are complete, i.e.~that their geometry is identical to the one that
would be found in a fiducial global mesh constructed in serial.

In order to robustly treat degenerate cases (for example when more than three
points lie on a common circle), we work with a computational volume that is
mapped to double precision numbers in the interval $[1,2[$. For IEEE
    arithmetic, the mantissa of these numbers effectively defines a one-to-one
    mapping of all representable floating point numbers in this range to the
    space of 53-bit integers. We then evaluate geometric predicates with fast
    ordinary double precision arithmetic, but always monitor the maximum
    round-off error. If the outcome of a predicate is not guaranteed to be
    correct as a result of round-off errors, we compute the predicate exactly
    with long-integer arithmetic based on the mantissas corresponding to the
    floating point numbers. We find that this approach is both robust and
    still quite fast.

Finally, we would like to mention that our moving-mesh approach can quite
easily be coupled to self-gravity using similar algorithms as are often
employed in particle-based SPH codes. In fact, in the {\small AREPO} code we
use a similar TreePM solver for gravity as employed in the {\small GADGET}
code \citep{Springel2005}. In our approach, the Voronoi cells are treated
effectively as point masses with a gravitational softening length set equal to
the fiducial radius of the cell, as estimated from its volume.  The tree-code
has the advantage of being highly efficient also for strongly clustered
particle configurations, and it can be easily adapted to individual timestep
integration.

\section{Illustrative test problems} \label{SecHydroTests}
 
We now discuss a number of simple test problems that show the performance of
the moving mesh approach and illustrate its specific strengths. In some cases,
we will compare directly to SPH simulations based on the same initial
conditions. Also, we discuss differences in the solutions when the mesh is
instead kept stationary, in which case our method behaves equivalently to a
standard Eulerian scheme with second-order accuracy in space and time.

\begin{figure}
\begin{center}
\resizebox{3.9cm}{!}{\includegraphics{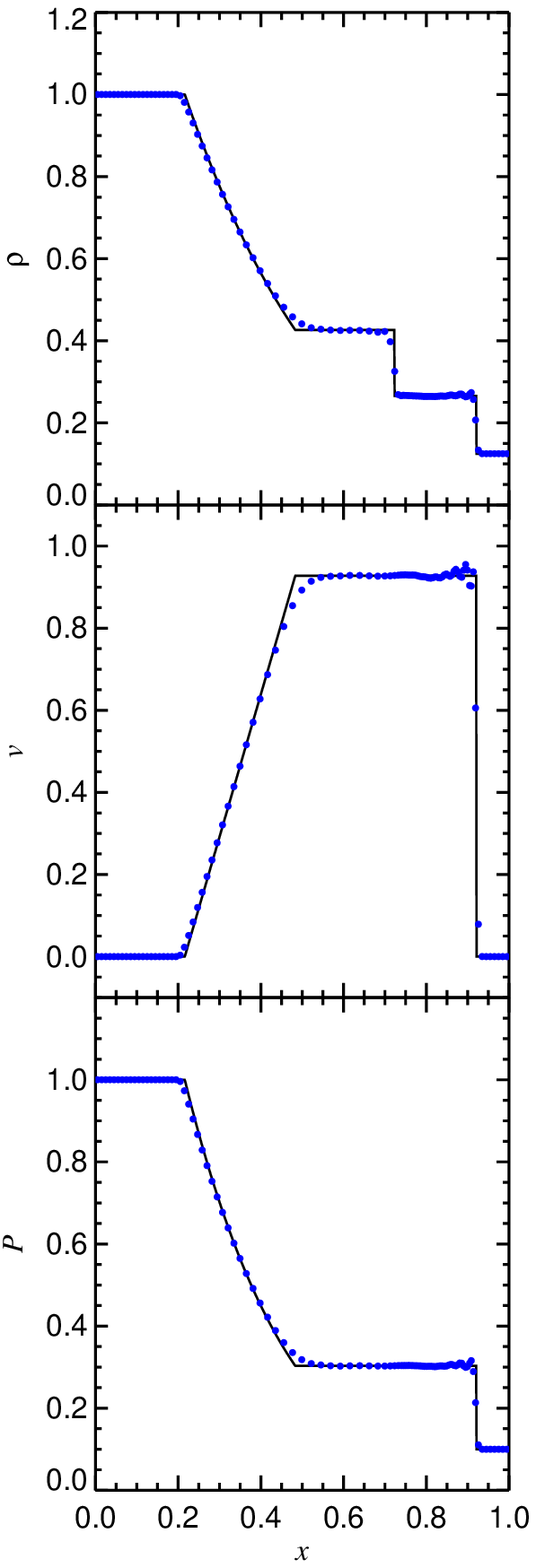}}%
\resizebox{3.9cm}{!}{\includegraphics{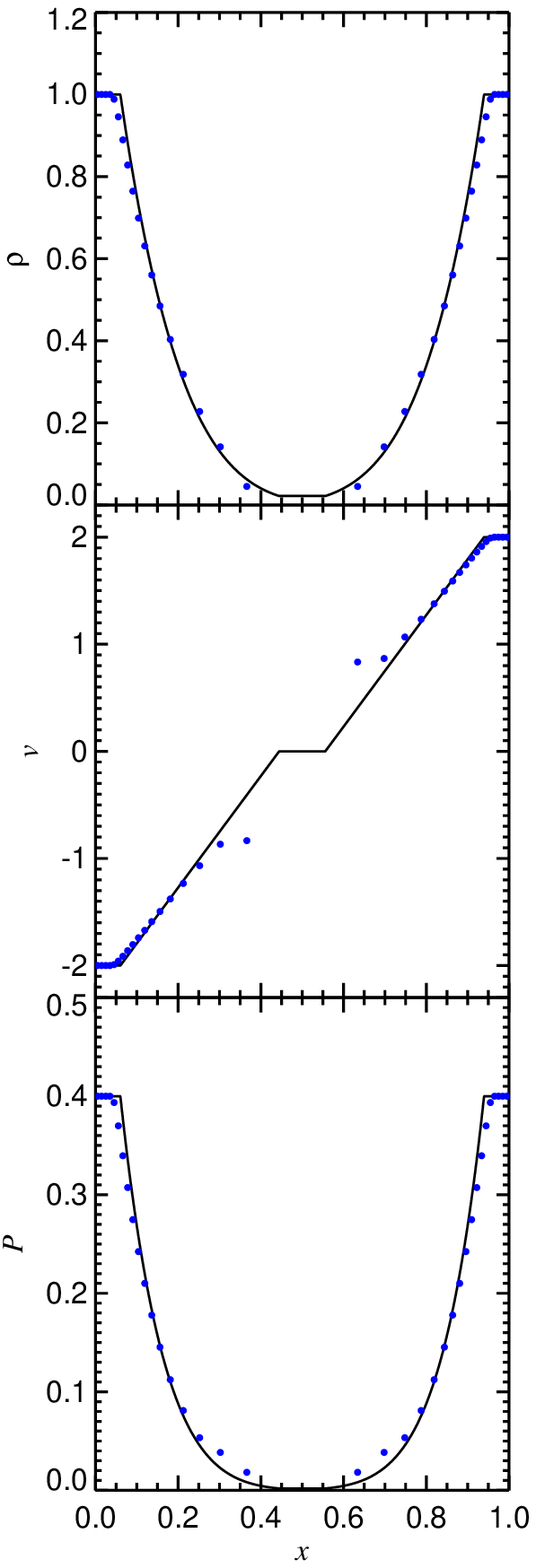}}%
\resizebox{3.9cm}{!}{\includegraphics{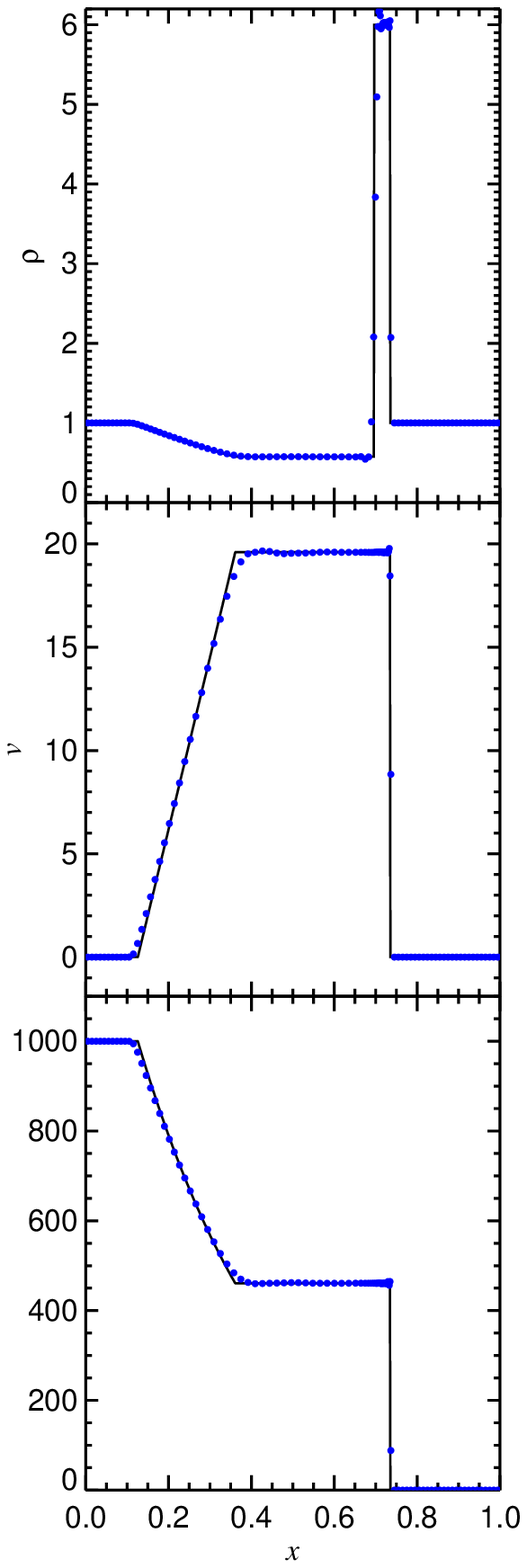}}
\vspace*{-0.5cm}
\end{center}
\caption{Different one-dimensional Riemann problems, calculated with a
  resolution of 100 points in the unit domain, for a gas with adiabatic index
  $\gamma=1.4$. The three columns show results for the initial conditions of
  the problems 1, 2 and 3 as specified in the text. Symbols represent the
  hydrodynamical quantities of the Voronoi cells, while the solid lines give
  the analytic solutions for density, velocity and pressure, from top to
  bottom.
 \label{Fig1DShocks}}
\end{figure}

\subsection{Riemann and Sod-shock problems}

Arguably the most important basic test problems of hydrodynamical codes
consist of one-dimensional Riemann problems. In the Riemann problem, two
piece-wise constant states, each characterized by density, pressure and
velocity, are brought into contact with each other, and their subsequent time
evolution is then followed. If the initial velocities are zero, one deals with
the special case of a Sod-shock problem.  The Riemann problem can be solved
analytically for an ideal gas, and the solution consists of a set of three
self-similar waves that emerge from the initial discontinuity. There is in
general one contact wave in the middle, sandwiched on either side by either a
shock or a rarefaction. The ability of a hydrodynamical method to accurately
treat different Riemann problems is fundamental for the ability of the scheme
to capture complex hydrodynamical phenomena.

In order to highlight the ability of our moving Voronoi mesh to accurately
represent Riemann problems we consider in Fig.~\ref{Fig1DShocks} the results
for three Riemann problems, simulated in 1D with 100 initially equally spaced
points in the unit domain. For problem one, the initial conditions are given
by $(\rho_1, P_1, v_1, \rho_2, P_2 , v_2)=(1.0, 1.0, 1.0, 0.125, 0.1, 0.0)$,
for problem 2 the corresponding values are $(1.0, 0.4, -2.0, 1.0, 0.4, 2.0)$,
and for problem 3 they are $(1.0, 1000.0, 0.0, 1.0, 0.01, 0.0)$. These values
describe the same problems as discussed in the book by \citet{Toro1997}, and
correspond to a moderate Sod shock tube, a strong double rarefaction, and a
very strong shock.

As we can see in Fig.~\ref{Fig1DShocks} from the comparison to the
analytic solution, the moving mesh approach captures the solutions of
these Riemann problems rather accurately. The contact discontinuities
and shocks are quite sharp, with only negligible post-shock
oscillations. The only significant error occurs in the nearly
evacuated region of the strong rarefaction, where the simulated
temperature is too high. However, this is a common error of Eulerian
codes when applied to this problem. We also see that the spatial
resolution varies at the end, since the points have moved with the
flow. In particular, the resolution has become quite low in the low
density region that develops in the middle of the double rarefaction,
while it has increased on the right hand side of the contact
discontinuities in the two Sod-shock problems. We note that the
accuracy with which the analytic solution is recovered is considerably
better than with SPH for the same initial conditions
\citep[see][]{Springel2010b}.

\subsection{Isentropic vortex convergence}

We next turn to a test of a non-trivial multi-dimensional fluid
problem with a smooth solution, the isentropic vortex problem
\citep{Yee2000,Calder2002}. This problem is particularly useful for
verifying whether our method does indeed show second-order
convergence, despite the presence of strong deformations and
topological changes of the mesh and the use of small velocity
components to keep the mesh nice and regular, as described
above. Previously, second-order convergence of the {\small AREPO} code
has only been explicitly demonstrated for 1D sound waves
\citep{Springel2010}, which is a comparatively simple problem where no
mesh twisting occurs. The 2D Gresho vortex test on the other hand
\citep{Gresho1990,Liska2003} shows only a convergence rate of $-1.4$
as a function of the number of cells per dimension
\citep{Springel2010}, both in our moving mesh code and other fixed
mesh codes that have second-order accuracy, like {\small ATHENA}
\citep{Stone2008}. However, this can be understood as a result of the
presence of discontinuities in the Gresho problem.

\begin{figure}
\begin{center}
\resizebox{6cm}{!}{\includegraphics{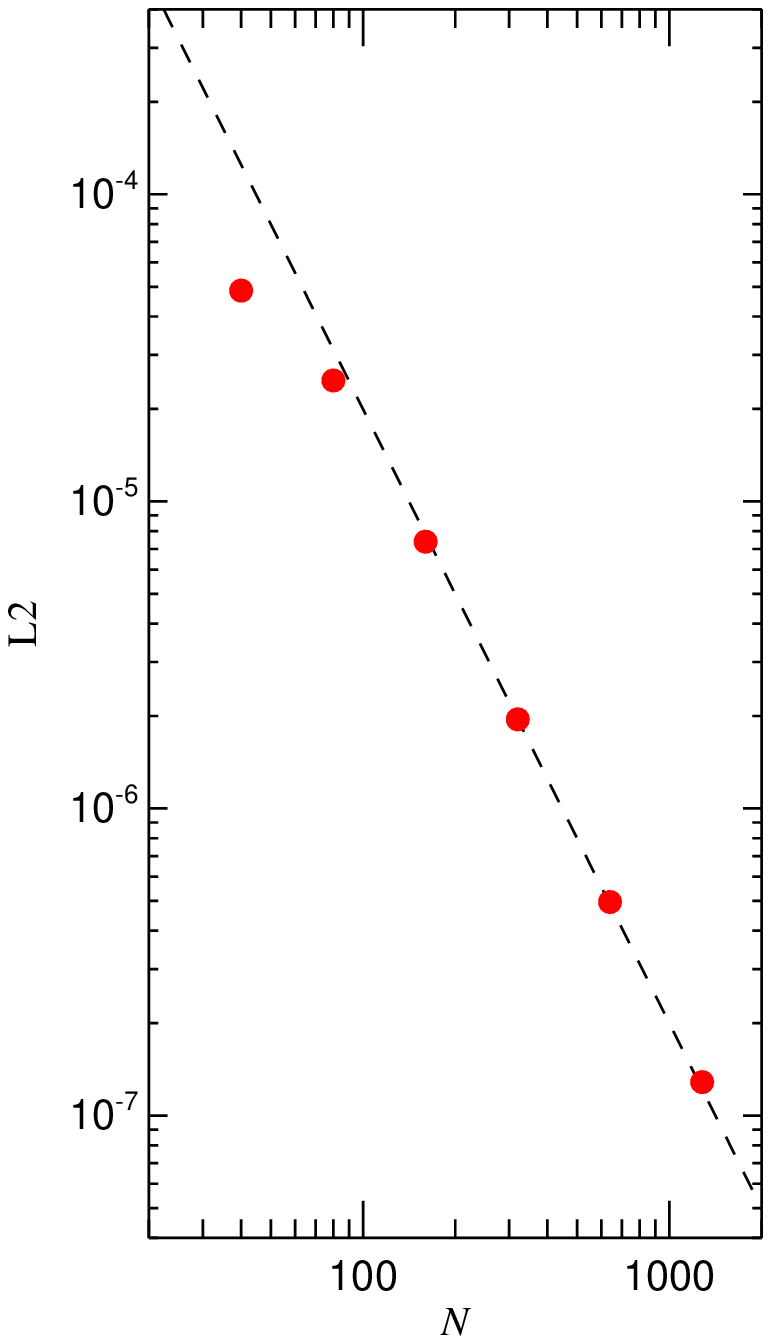}}%
\resizebox{5.9cm}{!}{\includegraphics{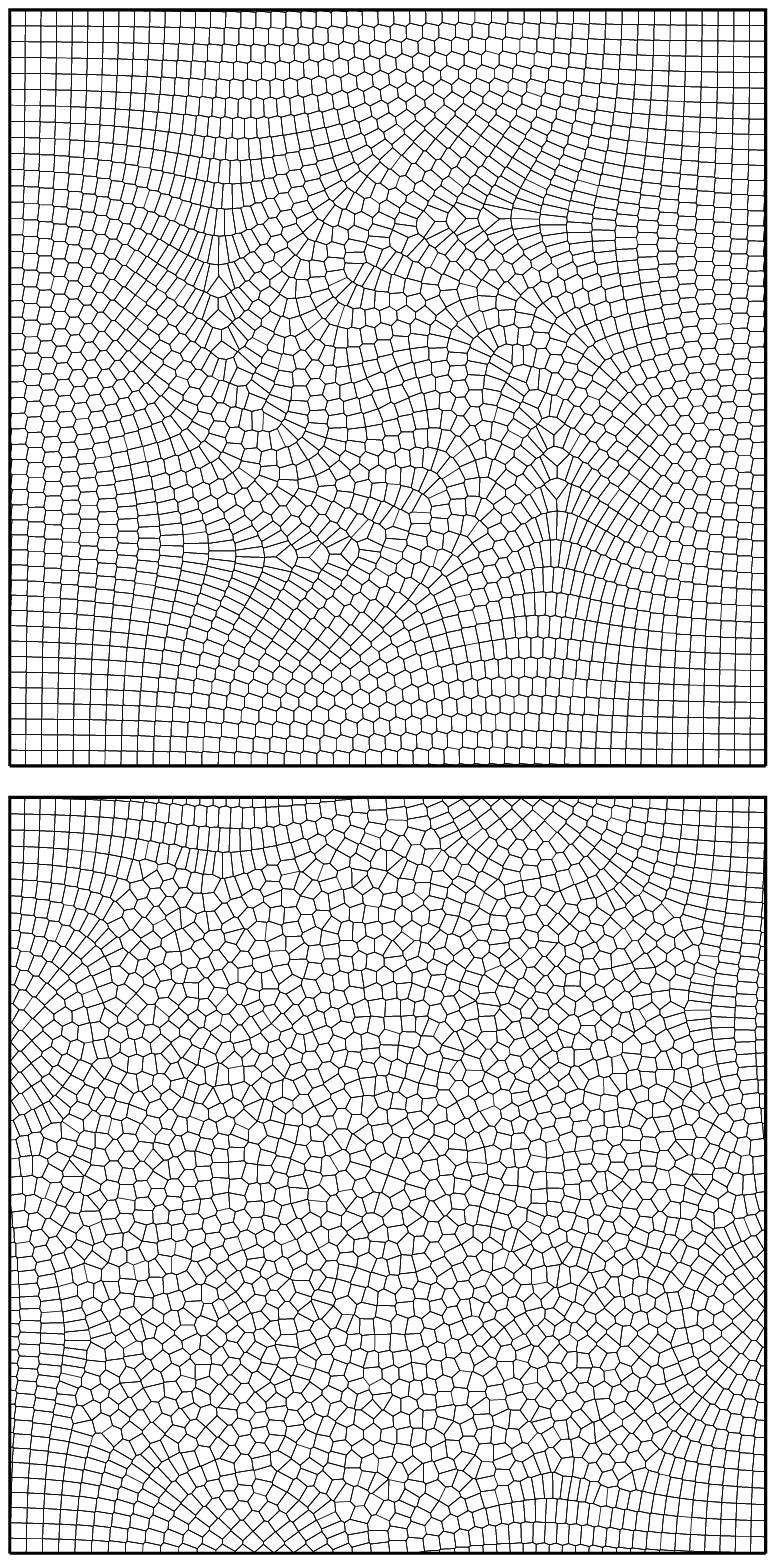}}%
\vspace*{-0.5cm}
\end{center}
\caption{L2 error norm at time $t=8.0$ as a function of resolution for
  the isentropic vortex test (left panel). The dashed line is a power
  law with ${\rm L2}\propto N^{-2}$. The panels on the right show the
  mesh in the region $[-4,4]\times [-4,4]$ around the center at times $t=1.5$
  (top) and $t=8.0$ (bottom), for the run at resolution $80\times 80$.
  \label{FigConvergence}}
\end{figure}

\citet{Yee2000} describe the setup of a perfectly smooth vortex, which
has an analytic, time-invariant solution.  To realize this problem, we
adopt a box of extension $[-5,5]^2$ in 2D, with periodic boundaries
everywhere. The initial distribution of mesh-generating points is
adopted as a regular Cartesian grid. The velocity field is specified
as
\begin{eqnarray}
v_x(x, y)  &= & -y \frac{\beta}{2 \pi} \exp\left(\frac{1-r^2}{2}\right)\\
v_y(x, y)  &= & x  \frac{\beta}{2 \pi}
\exp\left(\frac{1-r^2}{2}\right)
\end{eqnarray}
where $r^2 = x^2 + y^2$.
The density and thermal energy per unit mass are calculated from
\begin{equation}
T(x,y)  \equiv P / \rho = T_\infty  -    \frac{ (\gamma-1) \beta } {8 \gamma \pi^2}
\exp\left( 1-r^2 \right)
\end{equation}
as $\rho = T^{1/(\gamma-1)}$ and $u = T /(\gamma - 1)$.  For these
choices, the entropy $P/\rho^\gamma$ is exactly constant everywhere,
and the solution is time-independent. We adopt $\gamma = 1.4$, a vortex
strength $\beta = 5.0$, and $T_\infty =1$.  We realize the initial
conditions by integrating over the fields in each grid cell to obtain
the conserved variables, as in \citet{Calder2002}.  We then calculate
the evolution of the vortex with our moving mesh code for different
resolutions until time $t=8.0$, at which point the vortex has rotated
more than once, and the mesh has been thoroughly sheared in the
region of the vortex.

In Figure~\ref{FigConvergence}, we consider the L2-norm of the
numerically obtained density field at the final time relative to the
analytic solution, as a function of resolution. We use $N^2 = 40^2$,
$80^2$, $160^2$, $320^2$, $640^2$, and $1280^2$ initial mesh
cells. Reassuringly, the error declines accurately as a powerlaw, with
${\rm L2} \propto N^{-2}$, which is the expected convergence rate for
a second-order accurate scheme. This convergence rate has also been
reached by \citet{Calder2002} for the {\small FLASH} code, but unlike
for this code, the error in our approach is completely independent on
whether or not the vortex has an additional bulk velocity.  We note
that our result also disagrees with the conjecture that an additional
(second) mesh-construction per time step would be needed to reach this
convergence rate for multi-dimensional flow \citep{Duffell2011}.

\subsection{Rayleigh-Taylor instabilities}

In multi-dimensional flows, a further class of important hydrodynamical
phenomena besides acoustic waves and the non-linear waves related to Riemann
problems (shocks, contact discontinuities and rarefaction waves)
appears. These are so-called fluid instabilities, such as the Rayleigh-Taylor
or Kelvin-Helmholtz instabilities. They are highly important for producing
turbulence and for inducing mixing processes between different phases of
fluids.

\begin{figure}
\begin{center}
\resizebox{3.8cm}{!}{\includegraphics{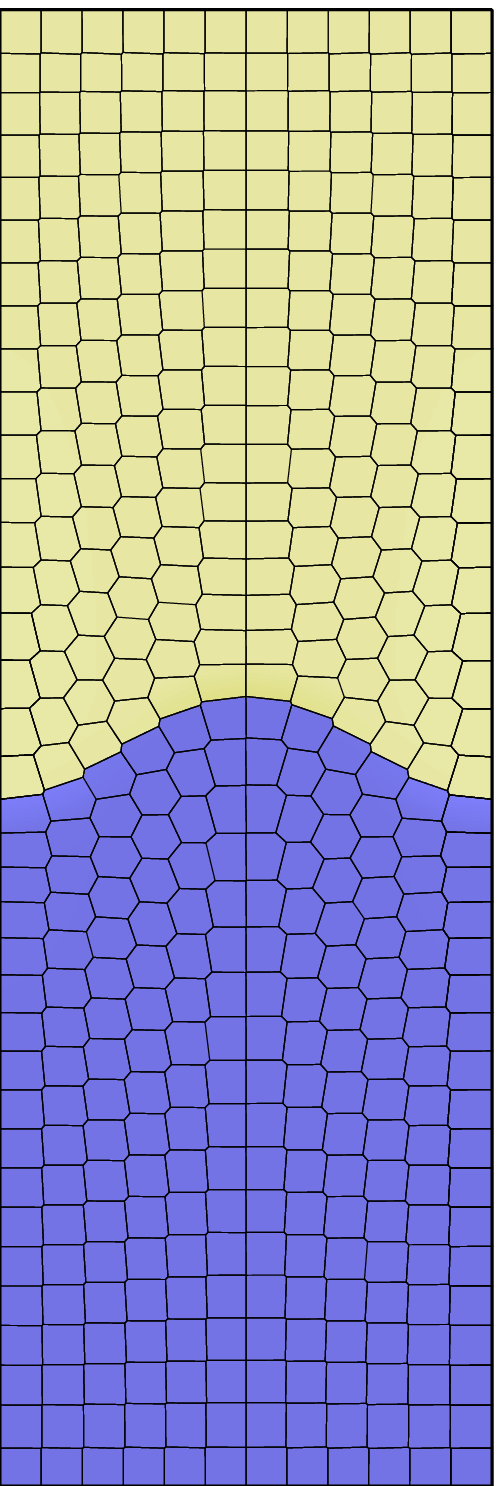}}\ %
\resizebox{3.8cm}{!}{\includegraphics{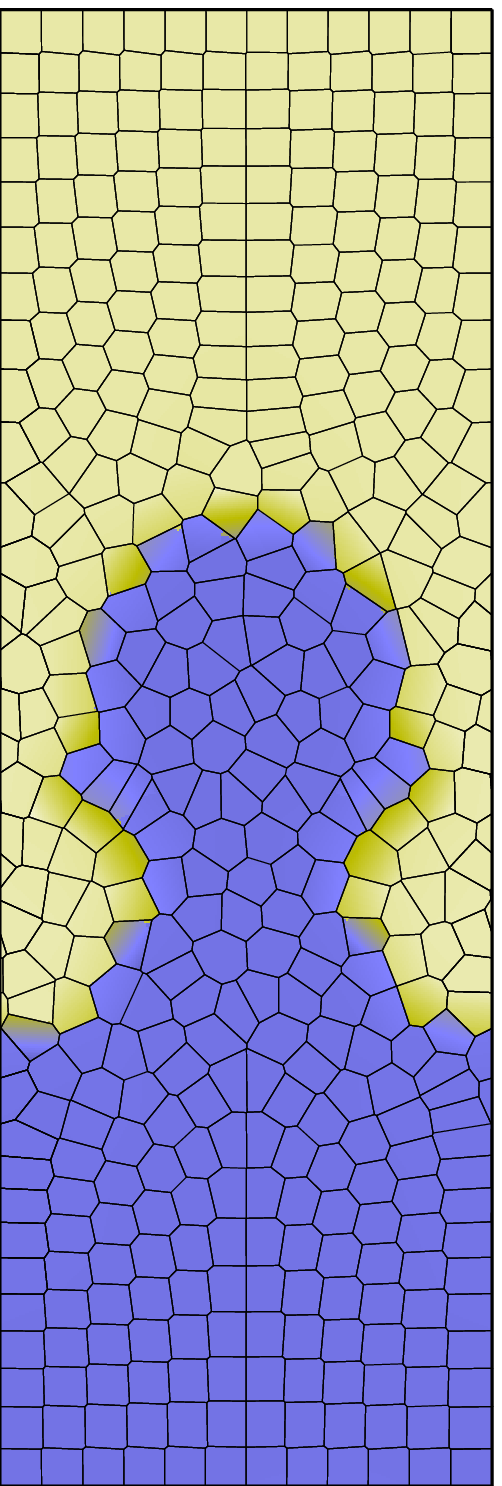}}\ %
\resizebox{3.8cm}{!}{\includegraphics{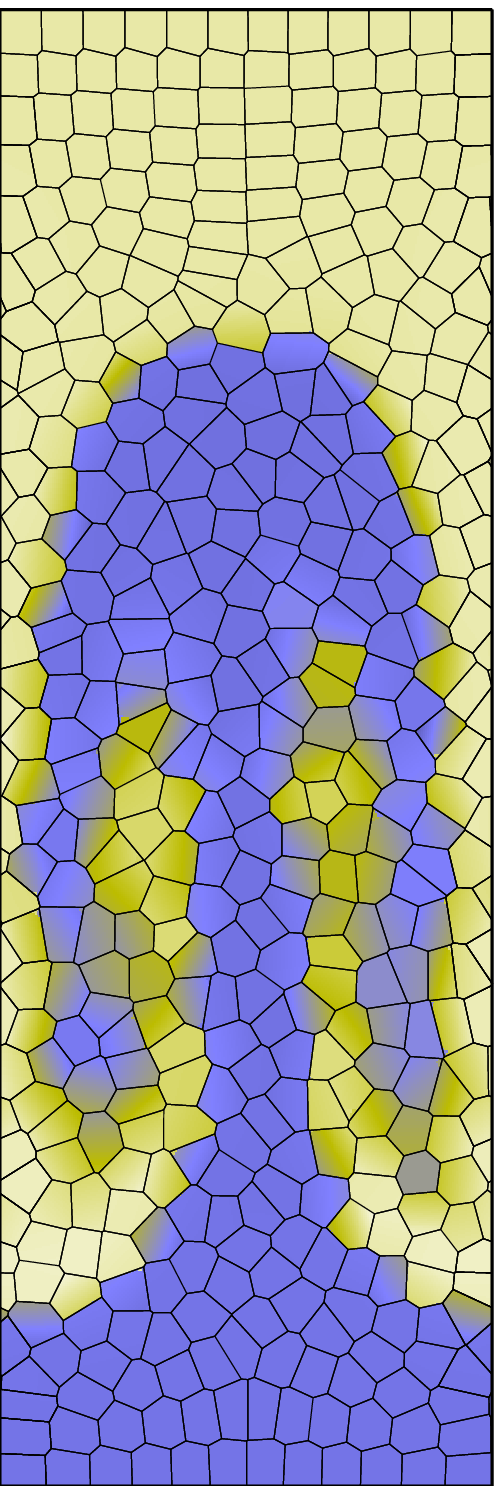}}
\vspace*{-0.2cm}
\end{center}
\caption{Rayleigh-Taylor instability calculated at low resolution with the
  moving-mesh approach. A denser fluid lies above a less dense fluid in an
  external gravitational field. The hydrostatic equilibrium of the initial
  state is buoyantly instable.  The three frames show the time evolution of the
  density field of the system at times $t=5.0$, $10.0$, and $15.0$, after a
  single mode has been perturbed to trigger the stability, as described in the
  text.
 \label{FigRTLowres}}
\end{figure}

The Rayleigh-Taylor instability can arise in stratified layers of gas in an
external gravitational field. If higher density gas lies on top of low-density
gas, the stratification is unstable to buoyancy forces, and characteristic
finger-like perturbations grow that will mix the fluids with time. To
illustrate this instability and simultaneously show the motion of the mesh in
our Voronoi based code, we illustrate in Figure~\ref{FigRTLowres} the
evolution of a single Rayleigh-Taylor mode, calculated at the deliberately low
resolution of $12\times 36$ cells. The simulation domain is two-dimensional,
with extension $[0.5,1.5]$ and periodic boundaries at the vertical boundaries,
and solid walls at the bottom and top. There is an external gravitational
field with acceleration $g=-0.1$, and the bottom and top halves of the box
are filled with gas of density $\rho=1$ and $\rho=2$, respectively. The
gravitational forces are balanced by an initial hydrostatic pressure profile
of the form $P(y) = P_0 + (y-0.75)\,g\,\rho(y)$ with $P_0$ and $\gamma=1.4$. To
seed the perturbation, one mode is excited with a small velocity perturbation
of the form $v_y(x,y)= w_0 [ 1 - \cos(4\pi x)][1-\cos(4\pi y/3)]$, where
$w_0=0.0025$.

\begin{figure}
\begin{center}
\resizebox{12cm}{!}{\includegraphics{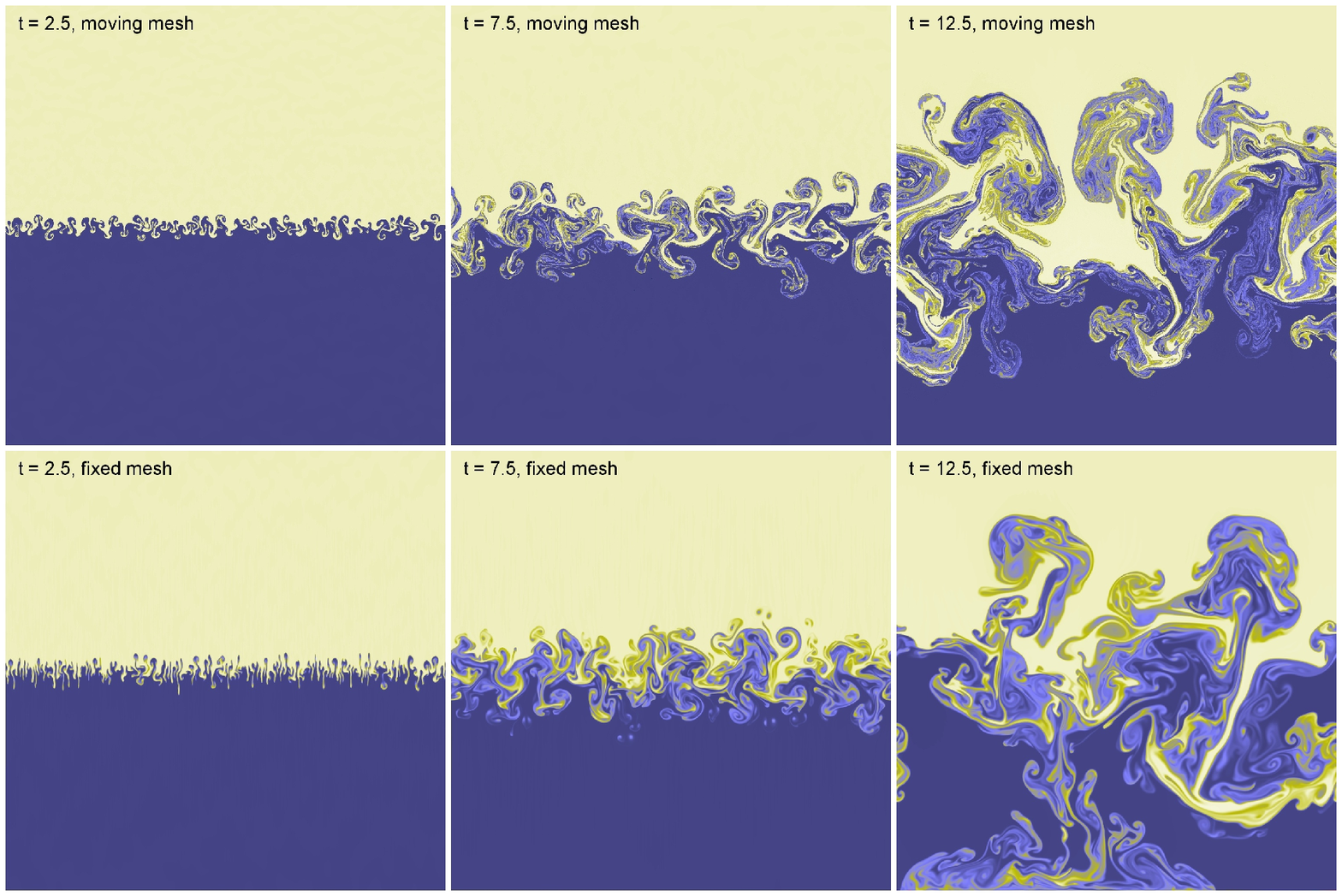}}%
\vspace*{-0.2cm}
\end{center}
\caption{Rayleigh-Taylor instability calculated at high resolution with
  $1024\times 1024$ points in the unit domain. The instability is here seeded
  by small random noise added to the velocity field. The top and bottom rows
  compare the time evolution for calculations with a moving and a stationary
  mesh, respectively.
 \label{FigRTHighres}}
\end{figure}

As can be clearly seen in the time evolution shown in
Figure~\ref{FigRTLowres}, the Rayleigh-Taylor instability is captured well by
the moving-mesh method even at this low resolution. What is particularly
interesting is that the sharp boundary between the phases can be maintained
for relatively long time during the early evolution of the instability, simply
because the contact discontinuity is not smeared out as it bends, thanks to
the mesh's ability to follow this motion in an approximately Lagrangian
fashion. A Eulerian approach with a stationary mesh on the other hand would
automatically wash out the boundary due to advection errors, involving some
spurious mixing of the fluids.

This fundamental improvement of the moving mesh code with respect to a
fixed mesh approach becomes clearer in Figure~\ref{FigRTHighres}. Here
we compare a high-resolution version of the Rayleigh-Taylor
instability between the moving-mesh approach and the same calculation
carried out with a stationary Cartesian mesh. Here $1024\times 1024$
cells have been used in the unit domain, $[-0.5,0.5]^2$, and the
instability was triggered by adding small random noise to the
$y$-velocity field, of the form $v_y(x,y)= A\, [ 1 + \cos(2\pi y)] /2$,
where $A$ is a random number in the interval $[-0.05, 0.05]$.  While
the instability shows a similar overall growth rate in both cases,
eventually leading to full turbulence in the box, there are also
striking differences. Whereas the calculation with the fixed mesh
produces a lot of intermediate density values due to the strong mixing
of the phases on small scales, the moving mesh approach maintains
finely stratified regions where different layers of the fluid phases
have been folded over each other. The contact discontinuities between
these layers can be kept sharp by the code even when they are moving
relative to the rest-frame of the box. We think that this behavior is
much more faithful to the underlying hydrodynamical flow. In the early
phase of the growth, it also appears as if small-scale RT fingers grow
somewhat too quickly in the Eulerian case as a result of grid
alignment effects.

\subsection{Kelvin-Helmholtz instabilities}

Another important fluid instability arises in shear flows, the so-called
Kelvin-Helmholtz (KH) instability. Whenever there is a discontinuity in the
shear velocity across a fluid interface, wave-like transverse perturbations
across the interface will grow into characteristic wave-like patterns. This
instability is ultimately behind the generation of waves on lakes and oceans
when wind streams over the water. The KH instability is ubiquitous in complex
flows and plays a prominent role in the generation of turbulence.

\begin{figure}
\centering \resizebox{5.8cm}{!}{\includegraphics{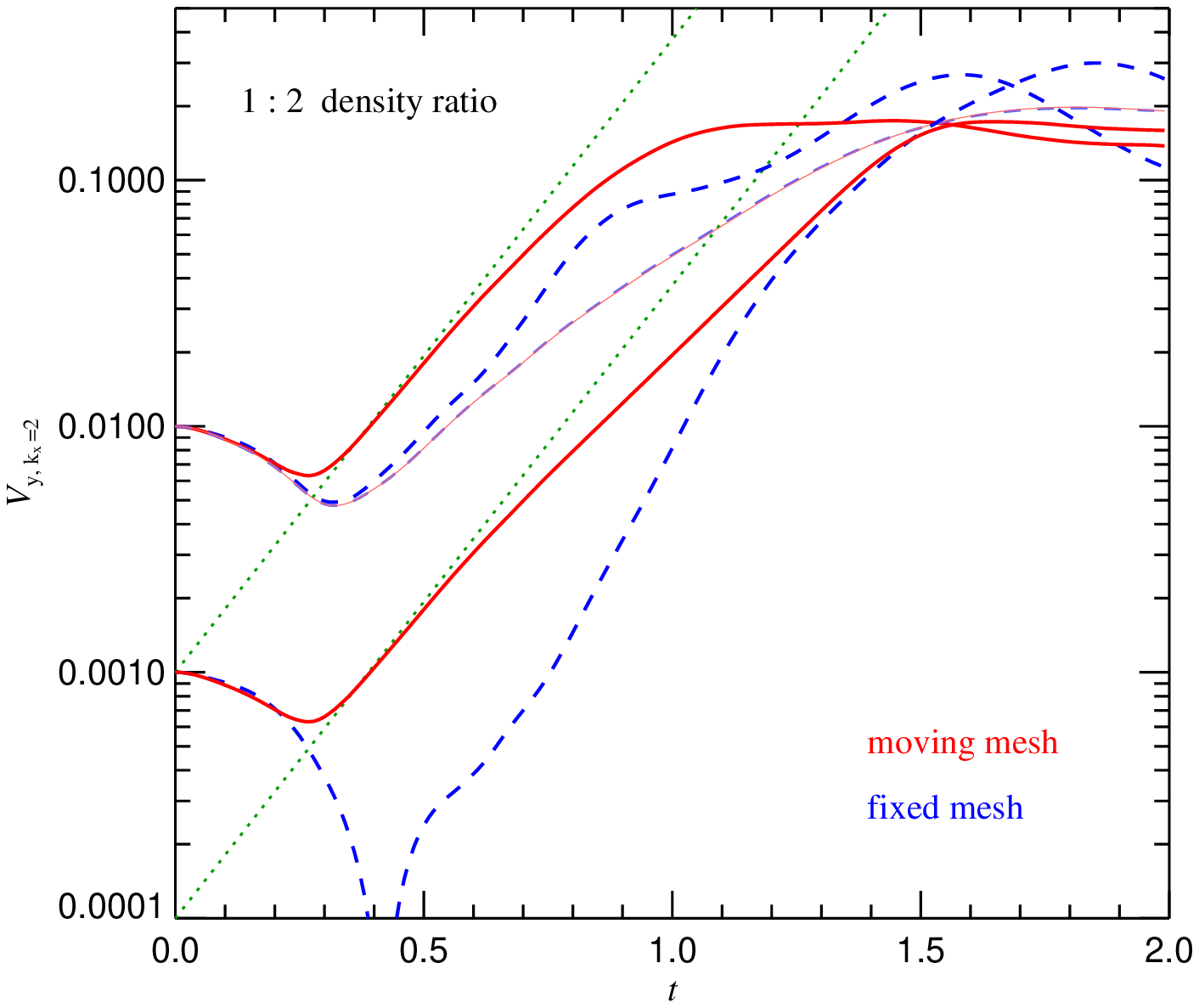}}%
\resizebox{5.8cm}{!}{\includegraphics{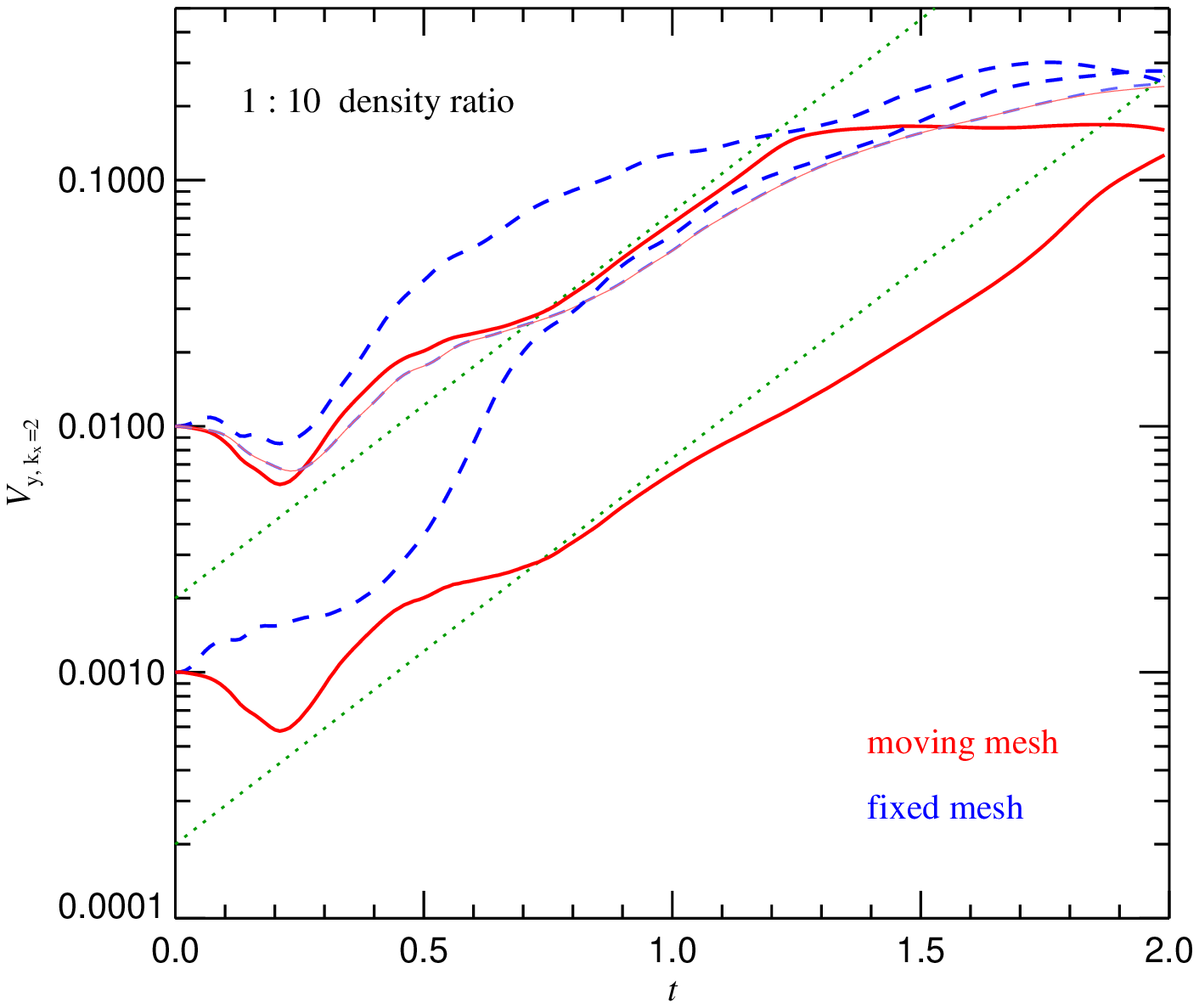}}
\caption{Growth rate of the KH instability for two different density ratios,
  $1:2$ (left panel) or $1:10$ (right panel).  The red lines show the results
  for the moving mesh code when initial conditions with a sharp density jump
  are used, either with an initial perturbation amplitude of $v_0=0.001$ or
  $v_0=0.01$. The dashed blue lines are the corresponding results for a
  stationary mesh. The dotted lines give the exponential growth expected from
  linear perturbation theory. Finally, the thin red and thin blue dashed lines
  are the results obtained when the initial discontinuity is washed out in the
  initial conditions. Time is given in units of the KH growth timescale
  $\tau_{\rm KH}$.
  \label{FigKHgrowthRate}}
\end{figure}

It has recently been found that the simulation of KH instabilities can be
quite problematic in SPH, with the growth being suppressed when the density
jump across the fluid discontinuity is large \citep{Agertz2007}.  This has
triggered a flurry of activity in the recent literature on SPH, trying to
improve on this behavior \citep{Price2007KH,Read2009,Hess2009,Junk2010,Abel2011}.

\begin{figure}
\centering
\resizebox{11.8cm}{!}{\includegraphics{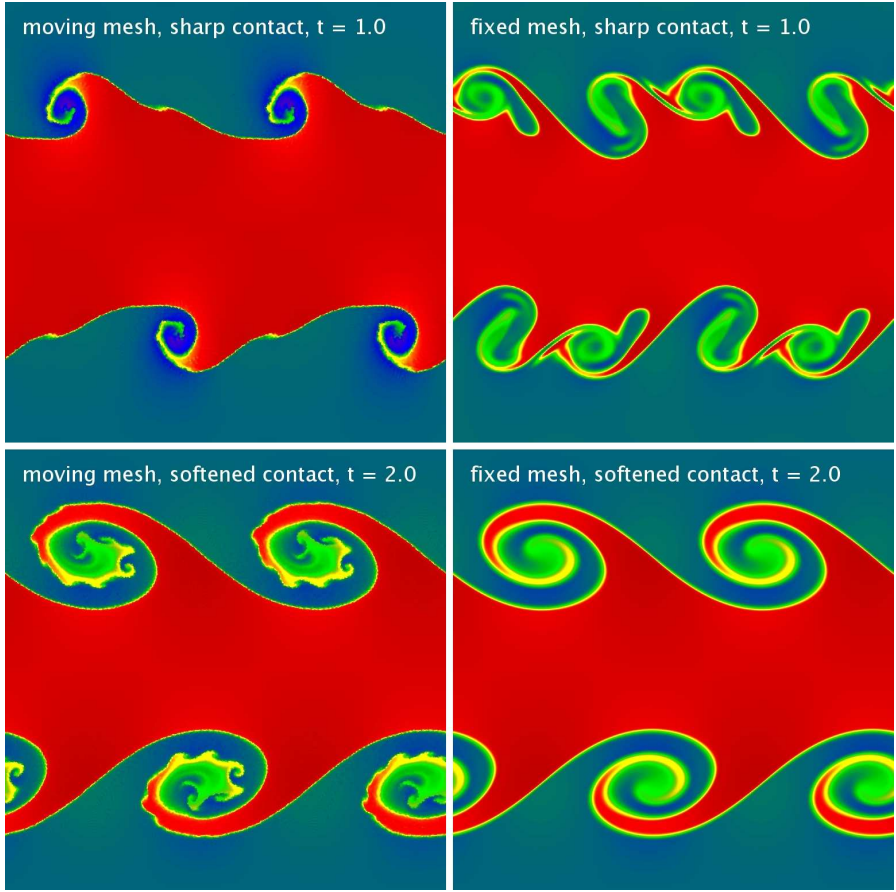}}
\caption{Kelvin Helmholtz instability computed with different initial
  conditions, and for a moving or a fixed mesh. In the panels on top, the
  initial contact discontinuity was sharp between adjacent cells. In contrast,
  in the bottom row it was smoothed out.
  \label{FigKHmaps}}
\end{figure}

We here show a basic KH test in a two-dimensional set-up, comparing our
moving-mesh approach against the traditional fixed-mesh approach. For
definiteness, we fill a box with periodic boundaries and unit length on a side
with gas of density $\rho_2=2$ in the horizontal middle stripe, and the rest
with density $\rho_1=1$. The middle region is moving to the right with
velocity $v_x=0.5$, the rest of the gas moves to the left with velocity
$v_x=-0.5$. The initial distribution of the mesh-generation points is a
Cartesian grid of resolution $256\times 256$.  We seed an initial perturbation
by adding an additional component 
\begin{equation}
v_y(x,y)= v_0 \, \sin( k\,x)
\end{equation}
to the velocity field, where $v_0$ is a small number. We choose $k=2 \times
(2\pi/L)$. Hence the Fourier spectrum of the $v_y$ field contains in the
beginning only the $k_x=2$ mode. The growth of this mode for $t>0$ can then be
conveniently measured through Fourier transforms of the velocity field.

In the left panel of Figure~\ref{FigKHgrowthRate}, a number of measurements of
the numerical KH growth rate for a density ratio of $1:2$ are summarized.  The
red lines show the result for our new moving-mesh method; the upper line is
for an initial perturbation amplitude of $v_0=0.01$, while the lower line is for
$v_0=0.001$. The blue dashed lines correspond in both cases to a fixed Cartesian
mesh of the same resolution. The linear theory
 growth rate $v_y
\propto \exp(t/\tau_{\rm KH})$ is shown as dotted lines, where
\begin{equation}
\tau_{\rm KH} = \frac{\rho_1 + \rho_2}{ |v_2-v_1|\,k \sqrt{\rho_1 \rho_2} }
\end{equation}
is the KH growth timescale for an inviscid gas.  Because the density is
initially not perturbed self-consistently with the velocity field, it takes
first a bit of time before the instability develops, but then the moving-mesh
solution follows the expected linear theory growth rate quite nicely for a
while. Eventually the growth slows down as the mode saturates and the
non-linear evolution of the KH instability ensues.

In the fixed-mesh case, the results are somewhat less clean and depend
more strongly on the initial perturbation amplitude. Visual inspection
of density maps during the time evolution reveals that not only the
excited mode starts to grow but also shorter wavelength modes. This is
not too surprising since small wavelength perturbations grow fastest
in the Kelvin-Helmholtz instability, and representing the initial
sharp density jump implicitly involves a spectrum of small waves. The
latter are prone to outgrow the larger-scale seed perturbation once
the discontinuity starts to be misaligned with the principal
coordinate axes, in which case additional small-scale perturbations
are seeded at mesh corners.

This effect can be repaired if the initial contact discontinuity is
washed out, as advocated by \citet{Robertson2010}. Then also in the
fixed-mesh case only the excited mode grows and a more stable result
is obtained. The latter is shown as thin dashed blue line in
Figure~\ref{FigKHgrowthRate}.  However, in this case one does not
reach the full growth rate that is expected analytically for the
(sharp) instability with this wavenumber. If one also applies the same
smoothing to the initial conditions of the moving-mesh run, one
obtains essentially the same result for the growth rate of the exited
mode, which is shown with a thin red line.  As soon as the initial
discontinuity is smooth enough to be resolved by several mesh cells,
it hence appears as if it would not make a difference whether one uses
a moving or a fixed-mesh. This is however not true. If one waits long
enough it is seen that the moving-mesh code resolves secondary KH
billows for which the fixed-mesh approach appears to be already too
diffusive. This can be seen in the density maps of
Figure~\ref{FigKHmaps}, where the bottom two panels compare the
density field of the KH test at time $t=2.0$ for the moving and the
fixed mesh approaches, using smoothed initial conditions. The top two
panels on the other hand give the same comparison when a sharp initial
discontinuity is used instead. In the latter case, it is clearly seen
that more `wrong' modes grow in the simulation with a stationary mesh,
because here a misalignment of the the sharp boundary with the mesh
triggers larger seed perturbations on small scales than for the moving
mesh. The question whether this initial condition is somehow `allowed'
or not \citep{Robertson2010} is moot in our view. Both codes are
started from identical initial conditions, and hence the comparison
tests how susceptible the codes are to the growth of numerically
seeded small-scale perturbations in this situation.

Finally, we have also considered the same test with a density jump of $1:10$,
which was realized by raising the density of the middle stripe to
$\rho_2=10$. The pressure was increased by a factor of 5 to $P=12.5$ in order
to ensure that the flow stays supersonic and the Mach number of the shear flow
is not much changed. The right panel in Figure~\ref{FigKHgrowthRate} shows the
corresponding results for the growth rate. Qualitatively, the results closely
follow those obtained for the $1:2$ density ratio, except that the fixed-mesh
results show a markedly too fast overall growth rate in this case.

\begin{figure}[t]
\centering
\resizebox{11.8cm}{!}{\includegraphics{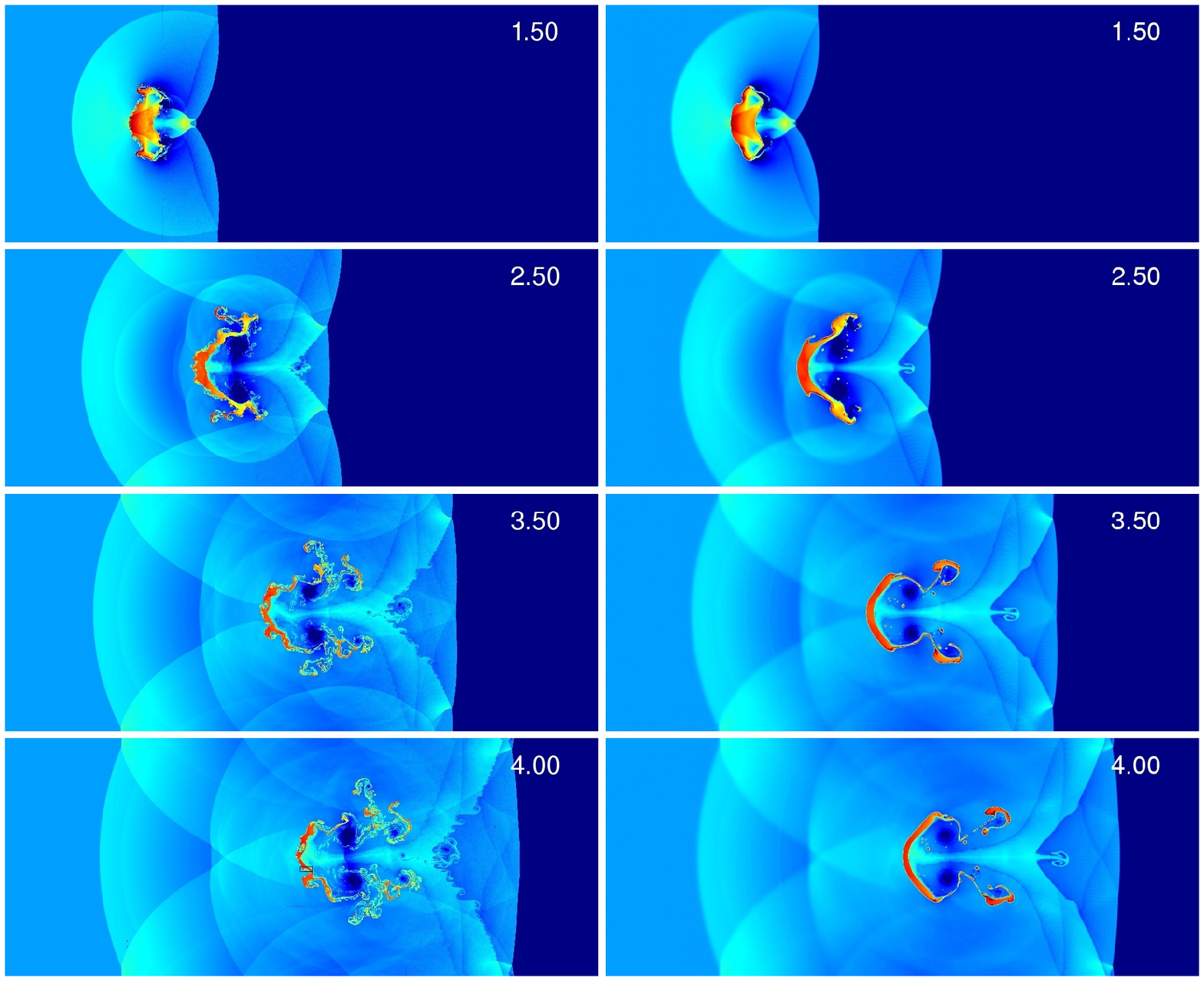}}\vspace{0.5cm}\\
\resizebox{5cm}{!}{\includegraphics{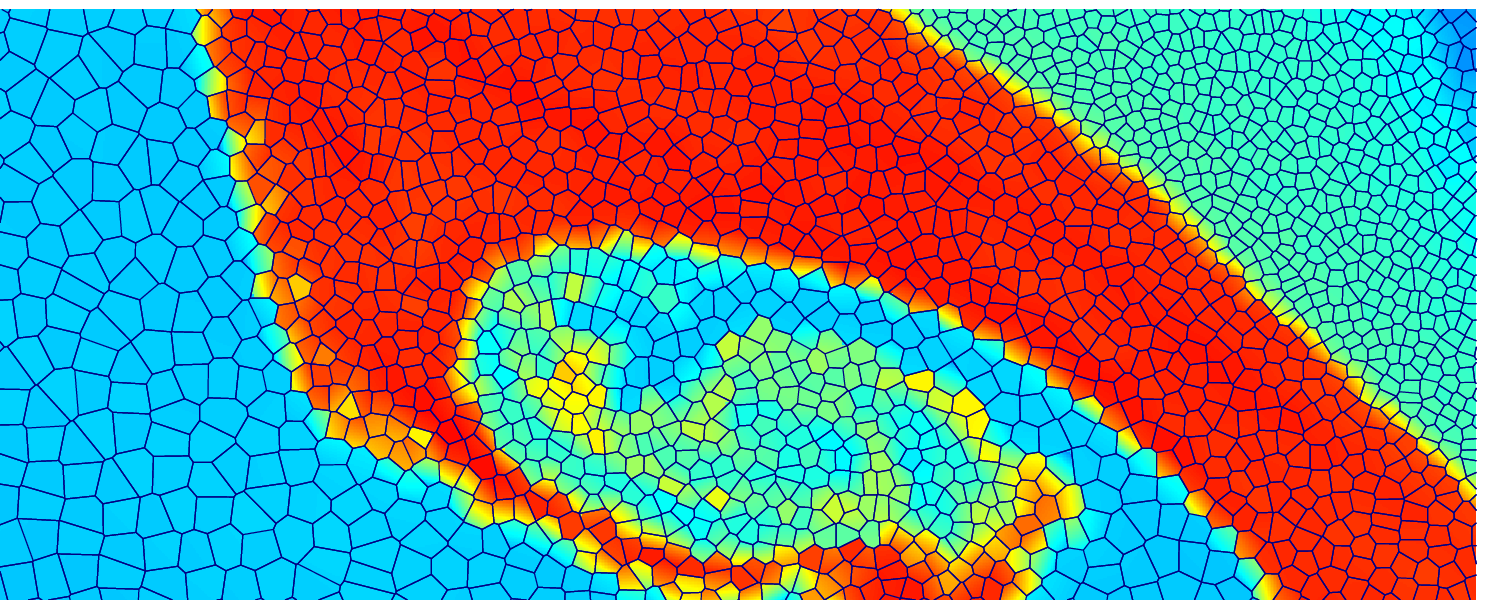}}\hspace*{1cm}%
\resizebox{5cm}{!}{\includegraphics{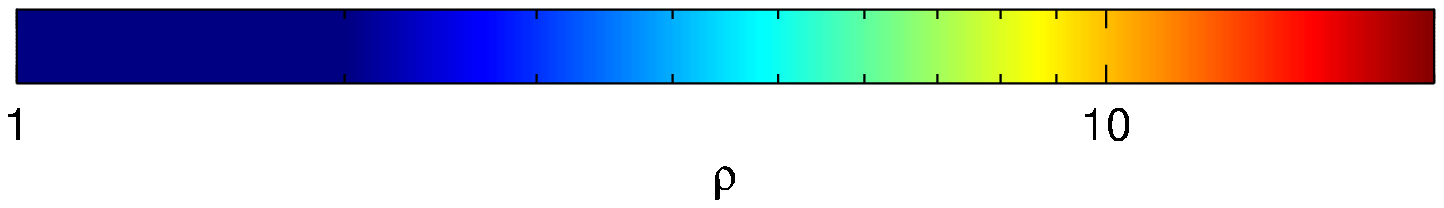}}
\caption{Comparison of the time evolution of the cloud disruption test in 2D
  with the moving mesh code (left column) and with SPH (right column). The
  panels show the density field at different times, as labeled. The same
  set-up and the identical initial conditions as in \citet{Springel2005} have
  been used. In the $t=4.0$ frame of the moving-mesh calculation, a small
  rectangle marks a region that is shown enlarged at the bottom left, with the
  Voronoi-mesh overlaid.
  \label{FigCloudDisruption}}
\end{figure}

\subsection{Shock-cloud interaction}

We finally consider two problems that involve the interaction of strong shocks
with fluid instabilities, which is important in many astrophysical
applications. First, we repeat a test presented in the {\small GADGET} code
paper \citep{Springel2005} as an advanced test of SPH. Here a strong shock
wave of Mach number 10 strikes an initially overdense cloud with density
$\rho=5$ that is embedded at pressure equilibrium in a tenuous hot phase of
density $\rho=1$ and pressure $P=1$ (with
$\gamma=5/3$). Figure~\ref{FigCloudDisruption} shows the time evolution of the
system in 2D, comparing the moving mesh results (left row) with the SPH result
obtained with the {\small GADGET} code for the identical initial conditions
(right row). As the shock strikes the cloud, it is compressed and
accelerated. A complicated system of reflected and interacting shocks
develops, and in the flow around the cloud, vortices are generated by the
baroclinic term. With time, these vortices tend to at least partially disrupt
the cloud.

In comparing the moving-mesh and the SPH results a number of interesting
observations can be made. First, the density field in the smooth regions is
noticeably noisy in the SPH calculation when compared with the moving-mesh
approach. Also, the shock waves are not as sharp and crisp as in the
Voronoi-based code, even though the global flow features are clearly very
similar in both cases. Arguably the most important difference is however that
the cloud is shredded much more in the moving-mesh simulation, while the SPH
result shows a large degree of coherence of the cloud debris. In fact, little
``droplets'' of dense gas remain that are eventually advected downstream in
the SPH calculation, showing no tendency to mix further with the background
gas. This is presumably related to a spurious surface tension effect in SPH
across contact discontinuities with large density jumps.

In the bottom left panel of the time-sequence shown in
Figure~\ref{FigCloudDisruption}, we have marked a small region with a black
rectangle. In order to illustrate the geometry of the Voronoi mesh in this
simulation, this region is shown enlarged at the bottom of
Fig.~\ref{FigCloudDisruption}, with the mesh overlaid. It can be seen that the
higher density region in the top right is populated with smaller Voronoi cells
than the lower density region at the bottom left, as a result of the
Lagrangian character of the scheme.

Finally, we turn to a related simulation problem in 3D, which has become known
as the `blob-test'. First carried out in \citet{Agertz2007}, this consists of
a three-dimensional overdense sphere that is put into a low-density background
gas that streams supersonically with respect to the cloud. We adopt the same
parameters as in \citet{Agertz2007}, and use the original mesh-based initial
conditions of this test as made available on the internet\footnote{They can be
  downloaded at http://www.astrosim.net}, at three different resolutions equal
to $32 \times 32\times 64$, $64 \times 64\times 128$, and $128 \times
128\times 256$. Similar to the two-dimensional shock-cloud interaction problem
discussed above, the supersonic head wind leads to the development of a
shear-flow over the surface of the cloud, which produces disrupting KH
instabilities. In \citet{Agertz2007} it was found that the SPH calculations
would only lead to an incomplete destruction of the cloud, while the
considered Eulerian mesh-code predicted a complete destruction of the cloud
after a relatively short time. The latter was measured in terms of the
mass-fraction of the original cloud that was still denser than $0.64$ times
the density of the cloud in the initial conditions and colder than $0.9$ times
the temperature of the background gas.

\begin{figure}[t]
\centering
\resizebox{10cm}{!}{\includegraphics{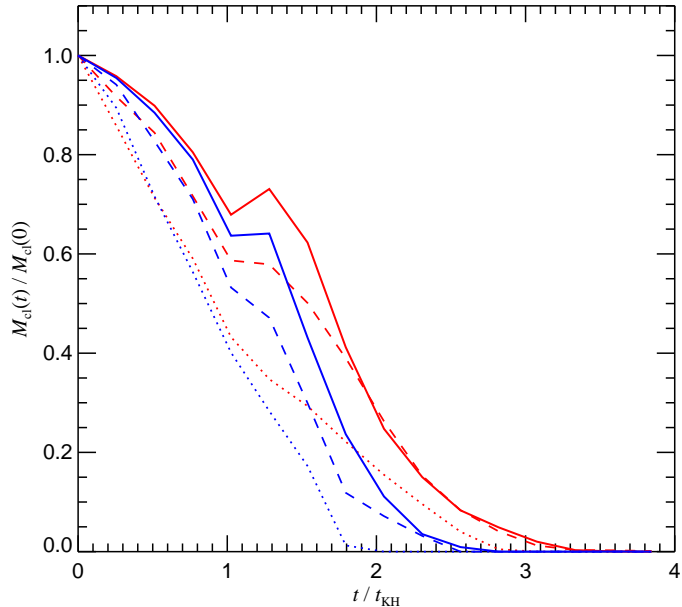}}%
\caption{The remaining mass fraction of a dense gas blob as a function of time
  when it is put into a supersonic head wind, corresponding to the
  three-dimensional `blob test' of \citet{Agertz2007}. The red solid, dashed,
  and dotted lines show our results for the moving Voronoi-mesh, with initial
  resolutions of $128 \times 128\times 256$, $64 \times 64\times 128$, and $32
  \times 32\times 64$ cells, respectively. The blue lines give the results
  when the mesh is kept stationary instead.
  \label{FigBlobMass}}
\end{figure}

In Figure~\ref{FigBlobMass}, we show our results for the remaining cloud mass
fraction as a function of time for three different resolutions, both for the
moving-mesh code and for the equivalent calculation with a stationary mesh.
We find that the moving-mesh approach appears to give converged results
already at lower resolution than the Eulerian approach. It also consistently
shows a slightly higher residual cloud mass fraction than the fixed mesh
calculation. Both effects can be understood as a result of the
Galilean-invariance and the considerably lower advection errors of the moving
mesh code. However, it is clear that both methods are qualitatively consistent
and predict a complete disruption of the cloud after a timescale of $t \sim
3\,\tau_{\rm KH}$. Because SPH gives a lower mass loss at late times and does
not produce a complete disruption of the cloud \citep{Agertz2007,Hess2009},
this reinforces the concern that gas stripping out of dense system can be
systematically underestimated in SPH.

\section{Discussion} \label{SecDiscussion}

We have described a novel hydrodynamical scheme on an unstructured mesh that
is constructed as the Voronoi tessellation of a finite set of mesh-generating
points.  The points are free to move during the time evolution, allowing the
mesh to seamlessly follow the flow and to change its spatial resolution fully
adaptively. Thanks to the mathematical properties of the Voronoi tessellation
there are no mesh-tangling or mesh-twisting effects since the motion of the
mesh-generating points induces a continuous deformation of the mesh, without
producing topological artefacts. 

Our approach represents a finite volume discretization of the Euler equations
on an unstructured mesh with second-order accuracy both in space and time,
without the need to invoke an artificial viscosity. Unlike ordinary Eulerian
codes, the new method is fully Galilean-invariant, which is a very substantial
advantage especially in simulations with large bulk flows. In particular, this
property implies high accuracy for contact discontinuities and leads to a
substantial reduction of advection errors when compared to traditional
Eulerian schemes. Indeed, these advantages of the moving-mesh approach can be
readily demonstrated with test problems involving fluid instabilities or
moving shock waves. The Voronoi-based approach also avoids preferred spatial
directions and offers flexibility in the treatment of boundary conditions. For
example, curved boundaries or moving interfaces can be readily implemented.

In this paper, we have not included a discussion of the technical
aspects of implementing the scheme in the new parallel cosmological
code {\small AREPO}, which is described in full detail in
\citet{Springel2010}. This implementation has already been applied to
timely problems in cosmological structure formation, such as galaxy
formation \citep{Vogelsberger2011} or the formation of the first stars
\citep{Greif2011a,Greif2011b}. Recently, important physics extensions
such as radiative transfer \citep{Petkova2011} and ideal
magnetohydrodynamics \citep{Pakmor2011} have been implemented in
{\small AREPO} as well. The code is hence becoming an increasingly
powerful alternative to more established simulation techniques in
astrophysics.

This is despite the considerable computational cost that the Voronoi
mesh construction entails, and despite the complicated bookkeeping
code that is required for the mesh management in parallel.  In 3D
hydrodynamics, our Voronoi code at present is about a factor of 2
slower for the same number of resolution elements than a SPH code (if
64 smoothing neighbours are used). Compared to a Eulerian fixed mesh
code, the speed difference is about a factor 3-4 (part of this
difference also stems from the about twice larger average number of
faces for our polyhedral cells compared with cubical cells in the
Cartesian case).  A further discussion of the speed difference and the
scalability of the code for large cosmological applications is given
in \citet{Vogelsberger2011}.  We note however that once self-gravity
is added, the relative speed differences are much reduced, as often a
sufficiently accurate calculation of gravity over a large dynamic
range is more expensive than the hydrodynamics itself. Further note
that our new technique reaches a given accuracy in many problems
already at a lower resolution than SPH and fixed mesh codes,
outweighing its higher complexity and making it hence also attractive
from the point of view of computational efficiency.

\index{paragraph}

\bibliographystyle{apj}
\bibliography{springel}

\begin{thebibliography}{61}
\expandafter\ifx\csname natexlab\endcsname\relax\def\natexlab#1{#1}\fi

\bibitem[{{Abel}(2011)}]{Abel2011}
{Abel}, T. 2011, \mnras, 413, 271

\bibitem[{{Agertz} {et~al.}(2007){Agertz}, {Moore}, {Stadel}, {Potter},
  {Miniati}, {Read}, {Mayer}, {Gawryszczak}, {Kravtsov}, {Nordlund}, {Pearce},
  {Quilis}, {Rudd}, {Springel}, {Stone}, {Tasker}, {Teyssier}, {Wadsley}, \&
  {Walder}}]{Agertz2007}
{Agertz}, O., {Moore}, B., {Stadel}, J., {Potter}, D., {Miniati}, F., {Read},
  J., {Mayer}, L., {Gawryszczak}, A., {Kravtsov}, A., {Nordlund}, {\AA}.,
  {Pearce}, F., {Quilis}, V., {Rudd}, D., {Springel}, V., {Stone}, J.,
  {Tasker}, E., {Teyssier}, R., {Wadsley}, J., \& {Walder}, R. 2007, \mnras,
  380, 963

\bibitem[{{Balsara}(2010)}]{Balsara2010}
{Balsara}, D.~S. 2010, Journal of Computational Physics, 229, 1970

\bibitem[{Barth \& Jesperson(1989)}]{Barth1989}
Barth, T.~J. \& Jesperson, D.~C. 1989, AIAA Paper, 89-0366

\bibitem[{{Braun} \& {Sambridge}(1995)}]{Braun1995}
{Braun}, J. \& {Sambridge}, M. 1995, \nat, 376, 655

\bibitem[{{Brio} {et~al.}(2001){Brio}, {Zakharian}, \& {Webb}}]{Brio2001}
{Brio}, M., {Zakharian}, A.~R., \& {Webb}, G.~M. 2001, Journal of Computational
  Physics, 167, 177

\bibitem[{{Calder} {et~al.}(2002){Calder}, {Fryxell}, {Plewa}, {Rosner},
  {Dursi}, {Weirs}, {Dupont}, {Robey}, {Kane}, {Remington}, {Drake}, {Dimonte},
  {Zingale}, {Timmes}, {Olson}, {Ricker}, {MacNeice}, \& {Tufo}}]{Calder2002}
{Calder}, A.~C., {Fryxell}, B., {Plewa}, T., {Rosner}, R., {Dursi}, L.~J.,
  {Weirs}, V.~G., {Dupont}, T., {Robey}, H.~F., {Kane}, J.~O., {Remington},
  B.~A., {Drake}, R.~P., {Dimonte}, G., {Zingale}, M., {Timmes}, F.~X.,
  {Olson}, K., {Ricker}, P., {MacNeice}, P., \& {Tufo}, H.~M. 2002, \apjs, 143,
  201

\bibitem[{Colella(1990)}]{Colella1990}
Colella, P. 1990, Journal of Computational Physics, 87, 171

\bibitem[{{Cunningham} {et~al.}(2009){Cunningham}, {Frank}, {Varni{\`e}re},
  {Mitran}, \& {Jones}}]{Cunningham2007}
{Cunningham}, A.~J., {Frank}, A., {Varni{\`e}re}, P., {Mitran}, S., \& {Jones},
  T.~W. 2009, \apjs, 182, 519

\bibitem[{{Duffell} \& {MacFadyen}(2011)}]{Duffell2011}
{Duffell}, P.~C. \& {MacFadyen}, A.~I. 2011, ArXiv e-prints, 1104.3562

\bibitem[{{Frenk} {et~al.}(1999){Frenk}, {White}, {Bode}, {Bond}, {Bryan},
  {Cen}, {Couchman}, {Evrard}, {Gnedin}, {Jenkins}, {Khokhlov}, {Klypin},
  {Navarro}, {Norman}, {Ostriker}, {Owen}, {Pearce}, {Pen}, {Steinmetz},
  {Thomas}, {Villumsen}, {Wadsley}, {Warren}, {Xu}, \& {Yepes}}]{Frenk1999}
{Frenk}, C.~S., {White}, S.~D.~M., {Bode}, P., {Bond}, J.~R., {Bryan}, G.~L.,
  {Cen}, R., {Couchman}, H.~M.~P., {Evrard}, A.~E., {Gnedin}, N., {Jenkins},
  A., {Khokhlov}, A.~M., {Klypin}, A., {Navarro}, J.~F., {Norman}, M.~L.,
  {Ostriker}, J.~P., {Owen}, J.~M., {Pearce}, F.~R., {Pen}, U.-L., {Steinmetz},
  M., {Thomas}, P.~A., {Villumsen}, J.~V., {Wadsley}, J.~W., {Warren}, M.~S.,
  {Xu}, G., \& {Yepes}, G. 1999, \apj, 525, 554

\bibitem[{{Fromang} {et~al.}(2006){Fromang}, {Hennebelle}, \&
  {Teyssier}}]{Fromang2006}
{Fromang}, S., {Hennebelle}, P., \& {Teyssier}, R. 2006, \aap, 457, 371

\bibitem[{{Gingold} \& {Monaghan}(1977)}]{Gingold1977}
{Gingold}, R.~A. \& {Monaghan}, J.~J. 1977, \mnras, 181, 375

\bibitem[{{Gnedin}(1995)}]{Gnedin1995}
{Gnedin}, N.~Y. 1995, \apjs, 97, 231

\bibitem[{{Greif} {et~al.}(2011{\natexlab{a}}){Greif}, {Springel}, {White},
  {Glover}, {Clark}, {Smith}, {Klessen}, \& {Bromm}}]{Greif2011a}
{Greif}, T.~H., {Springel}, V., {White}, S.~D.~M., {Glover}, S.~C.~O., {Clark},
  P.~C., {Smith}, R.~J., {Klessen}, R.~S., \& {Bromm}, V. 2011{\natexlab{a}},
  \apj, 737, 75

\bibitem[{{Greif} {et~al.}(2011{\natexlab{b}}){Greif}, {White}, {Klessen}, \&
  {Springel}}]{Greif2011b}
{Greif}, T.~H., {White}, S.~D.~M., {Klessen}, R.~S., \& {Springel}, V.
  2011{\natexlab{b}}, \apj, 736, 147

\bibitem[{{Gresho} \& Chan(1990)}]{Gresho1990}
{Gresho}, P.~M. \& Chan, S.~T. 1990, International Journal for Numerical
  Methods in Fluids, 11, 621

\bibitem[{Hassan {et~al.}(1998)Hassan, Probert, \& Morgan}]{Hassan1998}
Hassan, O., Probert, E.~J., \& Morgan, K. 1998, International Journal for
  Numerical Methods in Fluids, 27, 41

\bibitem[{{He{\ss}} \& {Springel}(2010)}]{Hess2009}
{He{\ss}}, S. \& {Springel}, V. 2010, \mnras, 406, 2289

\bibitem[{Hietel {et~al.}(2000)Hietel, Steiner, \& Struckmeier}]{Hietel2000}
Hietel, D., Steiner, K., \& Struckmeier, J. 2000, Mathematical Models and
  Methods in Applied Sciences, 10, 1363

\bibitem[{Junk(2002)}]{Junk2002}
Junk, M. 2002, in Lecture Notes in Computational Science and Engineering,
  Vol.~26, Meshfree Methods for Partial Differential Equations, ed. M.~Griebel,
  M.~A. Schweitzer, T.~J. Barth, M.~Griebel, D.~E. Keyes, R.~M. Nieminen,
  D.~Roose, \& T.~Schlick (Springer Berlin Heidelberg), 223--238

\bibitem[{{Junk} {et~al.}(2010){Junk}, {Walch}, {Heitsch}, {Burkert},
  {Wetzstein}, {Schartmann}, \& {Price}}]{Junk2010}
{Junk}, V., {Walch}, S., {Heitsch}, F., {Burkert}, A., {Wetzstein}, M.,
  {Schartmann}, M., \& {Price}, D. 2010, \mnras, 407, 1933

\bibitem[{{Katz} {et~al.}(1996){Katz}, {Weinberg}, \& {Hernquist}}]{Katz1996}
{Katz}, N., {Weinberg}, D.~H., \& {Hernquist}, L. 1996, \apjs, 105, 19

\bibitem[{{LeVeque}(2002)}]{LeVeque2002}
{LeVeque}, R.~J. 2002, Finite volume methods for hyperbolic systems (Cambridge
  University Press)

\bibitem[{Liska \& Wendroff(2003)}]{Liska2003}
Liska, R. \& Wendroff, B. 2003, SIAM J. Sci. Comput., 25, 995

\bibitem[{Lloyd(1982)}]{Lloyd1982}
Lloyd, S. 1982, IEEE Trans. Inform. Theory, 28, 129–137

\bibitem[{{Lucy}(1977)}]{Lucy1977}
{Lucy}, L.~B. 1977, \aj, 82, 1013

\bibitem[{Mavriplis(1997)}]{Mavripilis1997}
Mavriplis, D.~J. 1997, Annual Review of Fluid Mechanics, 29, 473

\bibitem[{{Mignone} {et~al.}(2007){Mignone}, {Bodo}, {Massaglia}, {Matsakos},
  {Tesileanu}, {Zanni}, \& {Ferrari}}]{Mignone2007}
{Mignone}, A., {Bodo}, G., {Massaglia}, S., {Matsakos}, T., {Tesileanu}, O.,
  {Zanni}, C., \& {Ferrari}, A. 2007, \apjs, 170, 228

\bibitem[{{Mitchell} {et~al.}(2009){Mitchell}, {McCarthy}, {Bower}, {Theuns},
  \& {Crain}}]{Mitchell2008}
{Mitchell}, N.~L., {McCarthy}, I.~G., {Bower}, R.~G., {Theuns}, T., \& {Crain},
  R.~A. 2009, \mnras, 395, 180

\bibitem[{{Monaghan}(1992)}]{Monaghan1992}
{Monaghan}, J.~J. 1992, \araa, 30, 543

\bibitem[{Okabe {et~al.}(2000)Okabe, Boots, Sugihara, \& Nok~Chiu}]{Okabe2000}
Okabe, A., Boots, B., Sugihara, K., \& Nok~Chiu, S. 2000, Spatial
  Tessellations, Concepts and Applications of Voronoi Diagrams (Chichester:
  John Wiley \& Sons Ltd)

\bibitem[{{Pakmor} {et~al.}(2011){Pakmor}, {Bauer}, \& {Springel}}]{Pakmor2011}
{Pakmor}, R., {Bauer}, A., \& {Springel}, V. 2011, MNRAS, accepted, ArXiv
  e-prints, 1108.1792

\bibitem[{{Pelupessy} {et~al.}(2003){Pelupessy}, {Schaap}, \& {van de
  Weygaert}}]{Pelupessy2003}
{Pelupessy}, F.~I., {Schaap}, W.~E., \& {van de Weygaert}, R. 2003, \aap, 403,
  389

\bibitem[{{Pen}(1998)}]{Pen1998}
{Pen}, U.-L. 1998, \apjs, 115, 19

\bibitem[{{Petkova} \& {Springel}(2011)}]{Petkova2011}
{Petkova}, M. \& {Springel}, V. 2011, \mnras, 415, 3731

\bibitem[{{Price}(2008)}]{Price2007KH}
{Price}, D.~J. 2008, Journal of Computational Physics, 227, 10040

\bibitem[{{Read} {et~al.}(2010){Read}, {Hayfield}, \& {Agertz}}]{Read2009}
{Read}, J.~I., {Hayfield}, T., \& {Agertz}, O. 2010, \mnras, 405, 1513

\bibitem[{{Robertson} {et~al.}(2010){Robertson}, {Kravtsov}, {Gnedin}, {Abel},
  \& {Rudd}}]{Robertson2010}
{Robertson}, B.~E., {Kravtsov}, A.~V., {Gnedin}, N.~Y., {Abel}, T., \& {Rudd},
  D.~H. 2010, \mnras, 401, 2463

\bibitem[{Sambridge {et~al.}(1995)Sambridge, Braun, \& McQueen}]{Sambridge1995}
Sambridge, M., Braun, J., \& McQueen, H. 1995, Geophysical Journal
  International, 122, 837

\bibitem[{{Schaap} \& {van de Weygaert}(2000)}]{Schaap2000}
{Schaap}, W.~E. \& {van de Weygaert}, R. 2000, \aap, 363, L29

\bibitem[{{Serrano} \& {Espa{\~n}ol}(2001)}]{Serrano2001}
{Serrano}, M. \& {Espa{\~n}ol}, P. 2001, Phys Rev E, 64, 046115

\bibitem[{{Serrano} {et~al.}(2005){Serrano}, {Espa{\~n}ol}, \&
  Zuniga}]{Serrano2005}
{Serrano}, M., {Espa{\~n}ol}, P., \& Zuniga, E. 2005, Journal of Statistical
  Physics, 121, 133

\bibitem[{{Springel}(2005)}]{Springel2005}
{Springel}, V. 2005, \mnras, 364, 1105

\bibitem[{{Springel}(2010{\natexlab{a}})}]{Springel2010}
---. 2010{\natexlab{a}}, \mnras, 401, 791

\bibitem[{{Springel}(2010{\natexlab{b}})}]{Springel2010b}
---. 2010{\natexlab{b}}, \araa, 48, 391

\bibitem[{{Springel} {et~al.}(2001){Springel}, {Yoshida}, \&
  {White}}]{Springel2001gadget}
{Springel}, V., {Yoshida}, N., \& {White}, S.~D.~M. 2001, New Astronomy, 6, 79

\bibitem[{{Stone} {et~al.}(2008){Stone}, {Gardiner}, {Teuben}, {Hawley}, \&
  {Simon}}]{Stone2008}
{Stone}, J.~M., {Gardiner}, T.~A., {Teuben}, P., {Hawley}, J.~F., \& {Simon},
  J.~B. 2008, \apjs, 178, 137

\bibitem[{{Tasker} {et~al.}(2008){Tasker}, {Brunino}, {Mitchell}, {Michielsen},
  {Hopton}, {Pearce}, {Bryan}, \& {Theuns}}]{Tasker2008}
{Tasker}, E.~J., {Brunino}, R., {Mitchell}, N.~L., {Michielsen}, D., {Hopton},
  S., {Pearce}, F.~R., {Bryan}, G.~L., \& {Theuns}, T. 2008, \mnras, 390, 1267

\bibitem[{{Toro}(1997)}]{Toro1997}
{Toro}, E. 1997, Riemann solvers and numerical methods for fluid dynamics
  (Springer)

\bibitem[{{Trac} \& {Pen}(2004)}]{Trac2004}
{Trac}, H. \& {Pen}, U.-L. 2004, New Astronomy, 9, 443

\bibitem[{{van de Weygaert}(1994)}]{Weygaert1994}
{van de Weygaert}, R. 1994, \aap, 283, 361

\bibitem[{{van de Weygaert} \& {Schaap}(2009)}]{vandeWeygaert2009}
{van de Weygaert}, R. \& {Schaap}, W. 2009, in Lecture Notes in Physics, Berlin
  Springer Verlag, Vol. 665, Data Analysis in Cosmology, ed.
  {V.~J.~Mart{\'{\i}}nez, E.~Saar, E.~Mart{\'{\i}}nez-Gonz{\'a}lez, \&
  M.-J.~Pons-Border{\'{\i}}a}, 291--413

\bibitem[{{van Leer}(1984)}]{Leer1984}
{van Leer}, B. 1984, SIAM J. Sci. Stat. Comput., 5, 1

\bibitem[{{van Leer}(2006)}]{Leer2006}
---. 2006, Communications in Computational Physics, 1, 192

\bibitem[{{Vogelsberger} {et~al.}(2011){Vogelsberger}, {Sijacki}, {Keres},
  {Springel}, \& {Hernquist}}]{Vogelsberger2011}
{Vogelsberger}, M., {Sijacki}, D., {Keres}, D., {Springel}, V., \& {Hernquist},
  L. 2011, ArXiv e-prints, 1109.1281

\bibitem[{{Wadsley} {et~al.}(2008){Wadsley}, {Veeravalli}, \&
  {Couchman}}]{Wadsley2008}
{Wadsley}, J.~W., {Veeravalli}, G., \& {Couchman}, H.~M.~P. 2008, \mnras, 387,
  427

\bibitem[{Wendroff(1999)}]{Wendroff1999}
Wendroff, B. 1999, Computers \& Mathematics with Applications, 38, 175

\bibitem[{{Whitehurst}(1995)}]{Whitehurst1995}
{Whitehurst}, R. 1995, \mnras, 277, 655

\bibitem[{{Xu}(1997)}]{Xu1997}
{Xu}, G. 1997, \mnras, 288, 903

\bibitem[{{Yee} {et~al.}(2000){Yee}, {Vinokur}, \& {Djomehri}}]{Yee2000}
{Yee}, H.~C., {Vinokur}, M., \& {Djomehri}, M.~J. 2000, Journal of
  Computational Physics, 162, 33

\end{thebibliography}

\printindex
\end{document}